\documentclass[12pt]{article}
\usepackage{comment}
\usepackage{xcolor}
\usepackage{commath}
\usepackage{soul}
\usepackage{CJKutf8}
\usepackage[normalem]{ulem}
\usepackage{graphicx}
\usepackage{ulem}
\usepackage{color}
\usepackage{amsmath,amssymb}
\usepackage{url}
\usepackage{hyperref}
\usepackage{scicite}

\usepackage{times}
\usepackage{txfonts}
\usepackage{booktabs}
\usepackage{subcaption}
\usepackage{threeparttable}

\def\kms{\mathrm{km\,s}^{-1}}
\def\mso{\,M_\odot}
\def\mi{M_\mathrm{1,ZAMS}}
\def\qi{q_\mathrm{ZAMS}}
\def\porbi{P_\mathrm{orb,ZAMS}}
\def\msoy{\, \mso~\mathrm{ yr}^{-1}}
\def\simgr{\mathrel{\hbox{\rlap{\hbox{\lower4pt\hbox{$\sim$}}}\hbox{$>$}}}}

\newcommand*\xbar[1]{%
   \hbox{%
     \vbox{%
       \hrule height 0.5pt %
       \kern0.5ex
       \hbox{%
         \kern-0.1em
         \ensuremath{#1}%
         \kern-0.1em
       }%
     }%
   }%
}

\newenvironment{sciabstract}{%
\begin{quote} \bf}
{\end{quote}}

\title{The formation of merging black holes via stable mass transfer in massive binary stars
}

{\centering
\author{
\begin{CJK*}{UTF8}{gbsn}
Xiao-Tian Xu (徐啸天)$^{1,2,3,*}$
\end{CJK*}
, Norbert Langer$^{3}$, Jakub Klencki$^{4,5}$,\\Chen Wang$^{2,6}$, Xiang-Dong Li$^{2,6}$\\
\\
\footnotesize{ $^{1}$Tsung-Dao Lee Institute, Shanghai Jiao-Tong University, 1 Lisuo Road, Shanghai 201210, People's Republic of China}\\
\footnotesize{ $^{2}$School of Astronomy and Space Science, Nanjing University, Nanjing 210093, People's Republic of China}\\
\footnotesize{$^{3}$Argelander-Institut f\"ur Astronomie, Universit\"at Bonn, Auf dem H\"ugel 71, 53121 Bonn, Germany}\\
\footnotesize{$^{4}$Max Planck Institute for Astrophysics, Karl-Schwarzschild-Strasse 1, 85748 Garching, Germany}\\
\footnotesize{$^{5}$European Southern Observatory, Karl-Schwarzschild-Str. 2, 85738 Garching bei München, Germany}\\
\footnotesize{$^{6}$Key Laboratory of Modern Astronomy and Astrophysics (Nanjing University), Ministry of Education,}\\
\footnotesize{ Nanjing 210023, People's Republic of China}
\\
\footnotesize{$^\ast$To whom correspondence should be addressed; E-mail:  xxu-tdli@sjtu.edu.cn, xxu@astro.uni-bonn.de}
}
}

\date{\today}

\begin{document}

\maketitle

\baselineskip24pt

\maketitle 

\bigskip

\textbf{Version Note:} this is pre-review version (old title: "Stable mass transfer in massive binaries leading to merging black holes"), and we refer to \url{https://www.nature.com/articles/s41550-026-03000-7} for the published version.

\begin{sciabstract}
The vast majority of massive binary systems in the universe is evidently unsuited to produce merging binary black holes. However, several narrow evolutionary paths of isolated massive binaries towards this goal have recently been identified. Due to the high degree of simplification and assumptions applied in previous modelling of these paths, conclusions remained vague so far. For one of these paths, the stable mass transfer channel, we now construct detailed binary evolution models which include internal differential rotation as well as mass and angular momentum transfer between the stars, all the way from the zero-age main sequence to the formation of the black holes, only skipping the rapid late burning stages. This allows us to follow the mass and chemical structure evolution of the mass accreting component, which turns out to have a key influence on the phase of reverse mass transfer, that allows the obtained black hole spins and mass ratios to naturally fall into the regime observed for the gravitational-wave source in the 10--25$\mso$ primary black hole mass range. As for this channel, also a large number of progenitor binaries are known, we conclude that it likely contributes to the observed population of gravitational wave sources. 
\end{sciabstract}

In the past ten years, about two hundred gravitational wave sources have been detected. The vast majority is caused by the merger of two stellar mass black holes (BHs) in binary black hole (BBH) systems \cite{LVK2025arXiv250818080T,LVK2025arXiv250818082T,LVK2025arXiv250818083T}. For BBHs with circular orbits to merge within the Hubble time ($\sim$14\,Gyr), the orbital separation of the binary at the formation time of the second BH needs to be smaller than a few Solar radii.
How BBHs that fulfil this condition can form remains highly debated. Primordial BHs of non-stellar origin have been suggested, but their binary properties remain unconstrained\cite{Belotsky2019}. For stellar-born BHs, merging BBHs may be produced by dynamical interactions in dense star clusters \cite{Rodriguez2019PhRvD.100d3027R}, in discs of active galactic nuclei  \cite{Samsing2022Natur.603..237S,Li2022MNRAS.517.1602L},
by  the evolution of massive triple stars \cite{Vigna-Gomez2021ApJ...907L..19V,Liu2018ApJ...863...68L},
or by isolated massive binary systems \cite{Belczynski2016Natur.534..512B,Mapelli2020,Mandel2022}.

The large fraction of massive stars found in close binary systems \cite{Sana2012} calls for investigating whether the undisturbed evolution of a subset of them can indeed lead to the formation of merging BBHs. Evidently, the vast majority of massive binary stars cannot, due to premature mergers, disruption caused by one of the stars exploding as supernova, or because the merger timescales of the produced BBHs exceed the Hubble time \cite{Gallegos-Garcia2021}. However, three possible pathway have been identified. The first one involves common envelope evolution (CEE) \cite{Belczynski2016Natur.534..512B,Mapelli2017,Kruckow2018,Shao2021ApJ...920...81S,Bavera2023NatAs...7.1090B}, which occurs when mass transfer in a binary system becomes unstable and the mass receiving star is engulfed in the envelope of the mass donor. 
While this process can lead to the required strong orbital shrinkage, most systems evolving along this path are expected to merge before the second BH has formed \cite{Belczynski2016,Mandel2022,Marchant2024ARA&A..62...21M}, and the conditions of entering and surviving CEE remain uncertain \cite{Ivanova2013,Ge2020,Marchant2021,Ropke2023LRCA....9....2R,Lau2025A&A...699A.274L,Chen2025arXiv251014173C}. 
A second proposed evolutionary path towards merging BBHs involves   tight massive binaries, in which both stellar components are spun up by tidal torques, and the induced rotational mixing keeps both stars chemically homogeneous, such that they never expand \cite{Yoon2005A&A...443..643Y,Woosley2006ApJ...637..914W}.  As this path requires both stars to be very massive and metal-poor \cite{Marchant2016,Eldridge2016MNRAS.462.3302E,Mandel2016MNRAS.458.2634M,Hastings2020A&A...641A..86H,Riley2022ApJS..258...34R}, it may best explain the most massive stellar BH mergers in the universe \cite{Popa2025arXiv250900154P}.

More recently, it has been suggested that merging BBHs can also be formed from binary stars which undergo stable mass transfer \cite{Marchant2021,Gallegos-Garcia2021,vanSon2022ApJ...940..184V,Olejak2024A&A...689A.305O,Picco2024A&A...681A..31P,Klencki2025arXiv250508860K}. 
In this path, the initially less massive star ends up more massive than the first formed black hole, and this allows the orbit to shrink upon mass transfer in the final stage.  Whether this tightens the orbit enough was shown to depend sensitively on the orbital configuration prior to the final mass transfer phase, and on the internal structure of the donor star \cite{Marchant2021, Klencki2025arXiv250508860K}, which was the mass gainer in the first mass transfer stage.

Previous efforts in modelling the stable mass transfer channel were often performed by using rapid binary evolution codes, as BSE\cite{Hurley2002}, StarTrack\cite{Belczynski2008ApJS..174..223B, Olejak2024A&A...689A.305O}, COMPAS\cite{Riley2022ApJS..258...34R, Broekgaarden2022ApJ...938...45B, Willcox2025arXiv251007573W}, COSMIC\cite{Breivik2020ApJ...898...71B}, and ComBinE\cite{Kruckow2018}. They rely on fitting formulae or tabulated data for massive single stars, and resolve each of the binary components by two grid points (a core, and an envelope). With simplifying assumption on the mass transfer physics, these codes perform a rudimentary bookkeeping of the mass and angular momentum budget of binaries stars throughout their evolution. Their high computational efficiency allows to explore large parameter variations, and led to vastly different predictions of the numbers and properties of merging BBHs \cite{Mandel2022}.
Detailed simulations towards merging BBHs, which solve the differential equations of stellar structure and evolution, often start after the formation of the first BH, with a zero-age main-sequence (ZAMS) star as companion (BH-ZAMS models)\cite{Marchant2021,Gallegos-Garcia2021,Shao2022ApJ...930...26S,Klencki2025arXiv250508860K}. 
The full evolutionary history of models in the stable mass transfer channel has been assessed using BPASS, which does not include differential rotation \cite{Eldridge2017PASA...34...58E,Briel2023MNRAS.520.5724B}, and MESA in the POSYDON framework \cite{Fragos2023ApJS..264...45F,Bavera2023NatAs...7.1090B}. Both efforts use the structure of a single star to approximate the main-sequence mass gainer after the first mass transfer phase\footnote{Mass transfer before the first BH formation is designated as the first mass transfer phase, and that after the BH formation as the second. }.

To overcome these simplifications and assumptions, we construct detailed massive binary models, which we evolve continuously from the ZAMS until the formation of the second BH with the MESA code version 8845 \cite{Paxton2011,Paxton2013,Paxton2015} (see Method for details). Our models fully resolve both stars simultaneously at all times. They include internal mixing processes that occur in accreting stars \cite{Braun1995A&A...297..483B,Renzo2021ApJ...923..277R}, internal differential rotation, tides, and mass and angular momentum transfer between the stars.  Since the progenitors of the observed merging BBHs were likely born in low-metallicity environment \cite{Belczynski2016Natur.534..512B,Klencki2025arXiv250508860K}, we compute our models with an initial chemical composition as observed in the interstellar medium of the Small Magellanic Cloud \cite{Brott2011}. To assess the role of the so far unexplored physics in the stable mass transfer channel, we compute a grid of models with a fixed primary star masses of $\mi = 31.6\mso$, for which we expect BH formation (Supplementary Section \ref{app:BH-formation}). We vary the initial mass ratio $\qi$ from 0.7 to 0.99, and explore initial orbital periods $\porbi$ below 4.5\,d (cf., Supplementary Section \ref{app:Pq-window}). In our fiducial models, 
we apply a rotation-limited mass transfer efficiency for non-degenerate mass gainers. We compute each of the two stars to core helium depletion, beyond which the mass, angular momentum and orbital properties are not expected to change considerably.
We assume BHs are formed without mass ejection (but see Supplementary  Section\,\ref{app:Xeff-mass-ejection}), and that accretion onto the BH is limited by the Eddington accretion rate. 

\section*{Results}

\subsection*{Example model}

Figure\,\ref{example_model} shows the evolution of the component masses and their internal structure for an example model, which starts out with two stars of 31.6$\mso$ and 28.4$\mso$ in a 2.1\,d orbit, and ends as a merging BBH. The initially more massive star expands first, triggering mass transfer when it is still core hydrogen burning \footnote{designated as Case\,A mass transfer, while mass transfer with a shell-hydrogen burning donor is called Case\,B \cite{Kippenhahn1967ZA.....65..251K}}.  
During the first mass transfer, the accretion efficiency is  near 90\%, such that the secondary star accretes  $\sim$15$\mso$ mass and grows to $\sim 43.3\mso$. This mass gain leads to the mixing of hydrogen into the core of the star, which makes it appear younger than its actual age (so-called rejuvenation; cf.\cite{Braun1995A&A...297..483B}). 
Furthermore, the accretion of helium-rich matter leads to an increase of the helium mass fraction in its envelope (Supplementary  Section \ref{app:kipp-single-and-rejuvenated}), making it more compact than a single star counterpart of the same mass. 

After the first mass transfer phase and the subsequent collapse of the stripped donor star, the binary models consists of a $13.2\mso$ BH and a $42\mso$ rejuvenated MS star with an orbital period of $\sim$6\,d (Supplementary Section \ref{app:kipp-CaseA-CaseB}). This configuration would undergo unstable Case\,A mass transfer and merger if the main sequence star would not have accreted helium-rich matter previously (\cite{Klencki2025arXiv250508860K}; Supplementary Section \ref{app:BH-ZAMS-1}). Our model performs stable Case\,B mass transfer here, as it remains rather compact during core hydrogen burning. We show below that this is important for the final spin parameter of the BH merger.

\subsection*{Zero-age main-sequence parameter space}

To identify the initial parameter space which leads to merging BBHs for our chosen primary mass, we compute 176 evolutionary models in the same way as our example system. Their evolutionary outcomes are presented in Fig.\,\ref{grid_outcomes}, which shows that only models with initial orbital periods below $\sim 4.2$\,d evolve into merging black holes, which restricts the first mass transfer to be of Case\,A. This Case\,A mass transfer occurs in three separate phases (cf., Fig.\,\ref{example_model}) and poses challenges to rapid binary evolution calculations. 
We find that, in our models, most of the initial parameter space of interacting binaries at this primary mass have a second mass transfer type of Case B (Case\,A-Case\,B model).
The reason is that most of the mass gainers in the first mass transfer undergo a more limited expansion during their core hydrogen burning than single stars of comparable mass, due to their helium-enriched envelops, and therefore do not fill their Roche volume until hydrogen shell burning expands their envelope. As indicated by the colour coding in Fig.\,\ref{grid_outcomes}, the shortest period ones form BBHs with merger times of 10\% of the Hubble time.

We also find Case\,A-Case\,A models to evolve into BBHs. However, their initial parameter space is more restricted ($\qi$: 0.7--0.8, and $\log\porbi$: 0.35--0.45), and their merger times are very close to the Hubble time, indicating that this part of the parameter space might be lost for BBH formation upon a slight variation of our uncertain physics input parameters.

\subsection*{Black hole-main-sequence star binaries}

When we compare the fate of our models just after the first BH formed with that of BH-ZAMS models (Fig.\,\ref{BH-MS}), we see first that in our models the range of MS star masses which can pair with $\sim 13\mso$ BHs is restricted to less than $\sim 44\mso$ by the mass budget in the first mass transfer. With random pairing, 13$\mso$ BH-ZAMS star systems with ZAMS star masses of up to $65\mso$ can form merging BBHs \cite{Klencki2025arXiv250508860K}. 
Furthermore, we see that in the majority of the parameter space of double black hole formation shown in Fig.\,\ref{BH-MS}, the BH-ZAMS models predict a Case\,A mass transfer, while our models undergo Case\,B mass transfer, for reasons discussed above. This is important because the evolutionary stage of the donor directly affects mass transfer stability \cite{Klencki2025arXiv250508860K}. Because of this shift from Case\,A to Case\,B for the second mass transfer,
our shortest period systems with the largest black hole companion masses (e.g., our example model) cover a region which is not reached by even the tightest BH-ZAMS models (cf.,\,Supplementary Section \ref{app:BH-ZAMS-1}), opening a new parameter space for the formation of merging BBHs. Notably, those models obtain merger delay times that are up to a factor of three shorter than that of the shortest delay time of the shown BH-ZAMS models ($\sim 0.34$ of the Hubble time).

\subsection*{Merging binary black holes}

Whether the second mass transfer is of Case\,A or Case\,B is not only important for the merger delay times, but also for the final black hole mass ratios and spins (cf., Fig.\,\ref{BH-BH}, and Supplementary Sections\,\ref{app:BHspin-compare} and \ref{app:tauM-chi}). The black hole mass ratios of our Case\,A-Case\,B models cluster at a value near 0.7. This is the natural outcome from the condition that the initial mass ratio of these systems is restricted to $\qi \simgr 0.8$, and a high mass transfer efficiency due to the small initial orbital separation required to obtain merging BBHs (see Supplementary Section\,\ref{app:qBBH} for more details). Notably, this makes the second formed BH more massive than the first formed one. 
Our Case\,A-Case\,A models, on the other hand, end up with black hole mass ratios near one, thus --- as the masses of the first born BH are all nearly the same --- with a lower mass of the second born BH. The reason is that they start out with smaller initial mass ratios 
due to the differences in the histories of mass accretion and internal mixing (Fig.\,\ref{grid_outcomes}). Additionally, their final Case\,A mass transfer leads to smaller helium core masses than a Case\,B mass transfer would for the same donor star \cite{Schurmann2024A&A...690A.282S}.

Figure\,\ref{BH-BH} shows that also the effective spin parameters of both groups of stars differ considerably. In our Case\,A-Case\,B models, both black holes rotate rather slowly, in line with previous finding for the SMT channel \cite{Bavera2021A&A...647A.153B,Olejak2021ApJ...921L...2O,Bavera2023NatAs...7.1090B}, which results in effective spin parameters of $0.1$--$0.25$. The spin parameters derived from our Case\,A-Case\,A models, on the other hand, cluster near a value of 0.6.
The reason is that in these models, the donor star in the second Case\,A mass transfer does not
expand after core hydrogen depletion due to its hydrogen-deficient envelope and therefore skips the second fast mass transfer phase associated with Case\,A (often called Case\,AB), thus losing less angular momentum (cf., Supplementary Sections\,\ref{app:BH-ZAMS-2}), which can result in a spin parameter as high as 0.8 for the second-born black hole (cf., Supplementary Section \ref{app:grid-details}). A similar evolution is also confirmed in more massive BH-ZAMS models (L. Ma, personal communication).
In Fig.\,\ref{BH-BH}, we see that the mass ratios of our Case\,A-Case\,B models correspond well with the gravitational wave data, while their spin parameters appear slightly too high. Our Case\,A-Case\,A models, on the other hand, appear at too high  mass ratios and spin parameters, in a region of the diagram where no BH mergers have been found (see Supplementary Section \ref{app:observed-BH-spin} for more details on the interpretation of observed BH spins), which may not be a significant tension. Given the large observational uncertainties, current gravitational wave data remain consistent with a considerable fraction of high-mass-ratio BH mergers \cite{LVK2025arXiv250818083T,Sridhar2025arXiv251122093S}.

\section*{Discussion and conclusion}

\subsection*{Model uncertainties}

While our simulations are affected by physics uncertainties, we find none of them is so severe that
the stable mass transfer channel towards merging BBHs is threatened not to work at all. Stellar rotation in our models is initialized by the condition of tidal locking at ZAMS. While the birth rotational velocities of massive stars are uncertain, our results are hardly affected by their choice, because the spin evolution of both stars is controlled by the binary interaction (Supplementary Section \ref{app:rotation}).
The efficiency of the first mass transfer phase in massive binaries is not well understood \cite{Langer2012}, but significant mass accretion is expected in close massive binaries \cite{Sen2022,Xu2025arXiv250323876X,Sen2025arXiv251115347S}. The accretor-spin controlled accretion-efficiency model we use, which generally results in very small efficiencies in wide binaries, even leads to nearly conservative evolution in our models (see also Supplementary Sections\,\ref{app:CaseA-CaseA} and \ref{app:kipp-CaseA-CaseB}). This is particularly true for our initially closest binary models, which lead to the shortest BBH merger delay times. The accretion efficiencies decreases to 70\%, and therefore becomes more uncertain, for our widest BBH merger progenitors. A decrease of the mass transfer efficiency might raise their merger times to above the Hubble time. We note that the stability criterion for mass transfer, and mass transfer efficiency of the second mass transfer are not very uncertain in detailed binary evolution models (cf., \cite{Klencki2025arXiv250508860K}).
The orbital properties of our predicted BH-MS binaries could be altered by BH birth kicks. However, in the considered BH mass range, large kicks are not expected \cite{Fryer2012ApJ...749...91F,Janka2013MNRAS.434.1355J} or empirically imposed \cite{Wong2012ApJ...747..111W,Mandel2016MNRAS.456..578M,Vigna-Gomez2024PhRvL.132s1403V,Willcox2025A&A...700A..59W}.

\subsection*{Progenitor observations of the stable mass transfer channel}

On their way to BBH mergers following the stable mass transfer channel, our models evolve through various distinct long-lasting evolutionary phases. Observed counterparts with roughly the right properties are observed for all of them. Close double O star binaries are found abundantly in the Milky Way and the Magellanic Clouds 
\cite{Sana2025NatAs...9.1337S}.
The first mass transfer phase produces a long-lived semi-detached evolution, whose counterparts, the massive Algol binaries, are also common \cite{Sen2022}. Thereafter, the  envelope-stripped mass donor would appear as a Wolf-Rayet star, corresponding to the observed short-period Wolf-Rayet+O-star binaries in the Milky Way \cite{Dsilva2020,Dsilva2022,Dsilva2023} and the Magellanic Clouds \cite{Shenar2016,Shenar2018,Schootemeijer2018,Pauli2023}, which we expect to be progenitors of BH+O-star systems,
like Cygnus X-1 \cite{Miller-Jones2021}, LMC X-1 \cite{Orosz2009}, and M33 X-7 \cite{Ramachandran2022}.
Even for the following BH-Wolf-Rayet-star stage, three candidate systems of Wolf-Rayet star-BH binaries have been identified, Cygnus X-3 \cite{Zdziarski2013}, NGC300 X-1 \cite{Binder2021}, and IC10 X-1 \cite{Laycock2015}.

\subsection*{Conclusion}

Detailed binary evolution simulations from the birth of the binary to the formation of the second BH, together with observed counterparts for all key evolutionary stages, establish the stable mass transfer channel as a robust contributor to the observed gravitational wave events. This is supported by the mass ratios and spin parameters produced by the models with relatively short merger delay times ($\sim 2\,$Gyr).  Our models show that it is imperative to compute the detailed evolution without interruption, in order to obtain the correct chemical structure of the mass gainer after the first mass transfer phase, which was not done previously. It changes the mass transfer history by allowing for tighter orbits in the second mass transfer, shorter merger times, smaller spins parameters and BH mass ratios. 

A fraction of our models performs the second mass transfer while the donor star undergoes core hydrogen burning, which produces spin parameters and BH mass ratios which appear incompatible with the gravitational wave data. At the same time, all of these models obtain merger delay times close to the age of the universe. It remains to be investigated whether their present day merger rate would be too small to matter, and whether small changes in the adopted physics parameter can remedy the issue.  

While it may remain challenging to capture the structure of mass gainers in rapid binary evolution models, MESA contains all relevant physics and allows to compute a large number of models.
Our experiment needs to be repeated at different primary masses to explore the mass dependence of the stable mass transfer channel. Physics parameters need to be varied in order to obtain BBH merger rate estimates and their uncertainties. Such model grids might also be pivotal in the interpretation of the structure in the observed distributions of the properties of the merging BBHs \cite{Antonini2025arXiv250904637A,Tong2025arXiv251105316T,Banagiri2025arXiv250915646B,Sridhar2025arXiv251122093S}, which start to emerge as the gravitational wave observatories keep finding BBH mergers at an increased pace \cite{LVK2025arXiv250818080T}.

\begin{figure*}[!htbp]
    \centering
    \includegraphics[width=\linewidth]{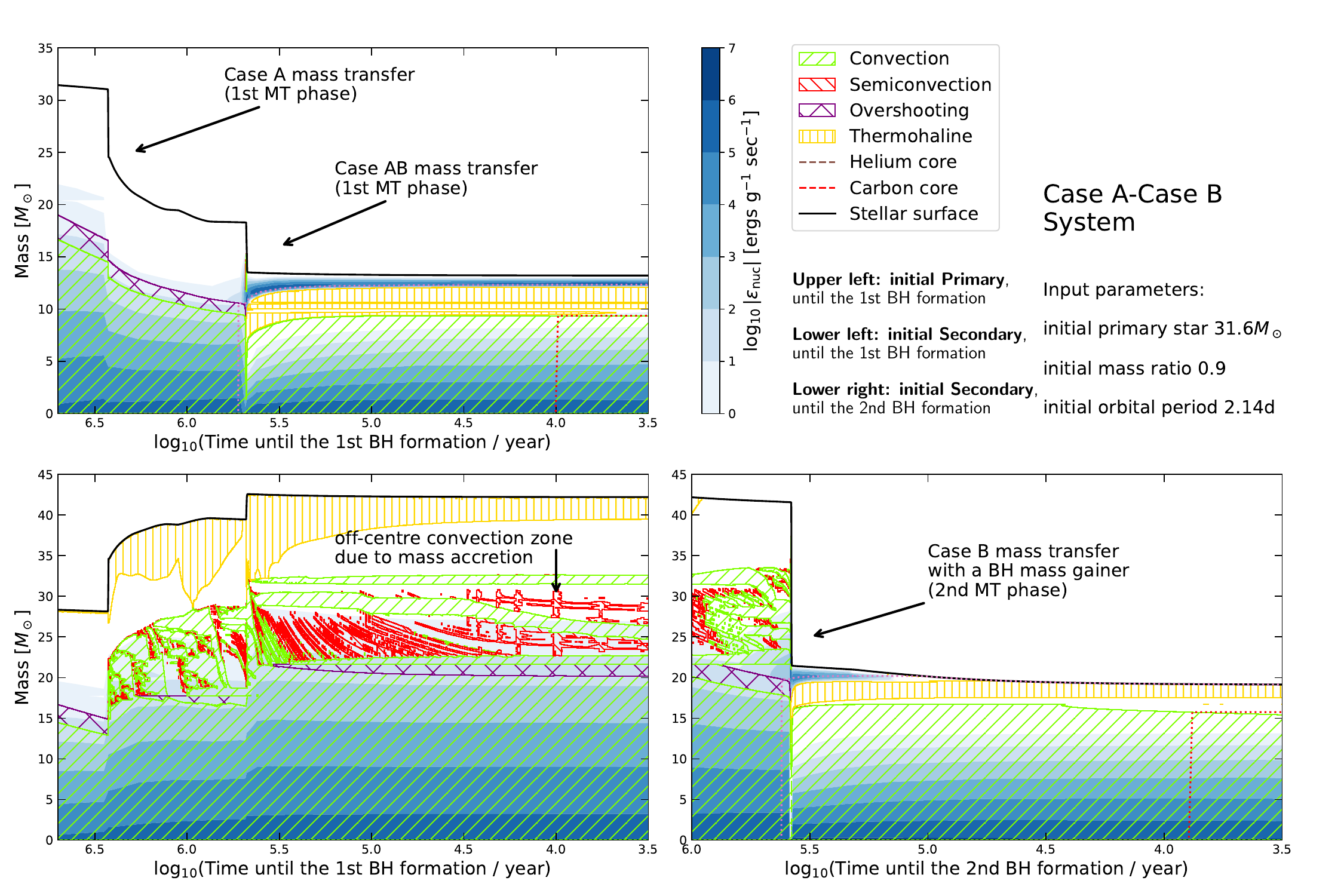}
    \caption{\textbf{Evolution of stellar structure (Kippenhahn diagrams) in the example model.} In each panel, the X-axis shows the time until the first or second BH formation event, and Y-axis represent the internal mass coordinate of the depicted star, from centre (0) to surface (black solid line). \textbf{Top left}: evolution of the initially more massive star from zero age until it forms a BH. \textbf{Bottom left}: evolution of the initially less massive star over the same time period as in the top panel, showing the rich history of internal chemical mixing induced by mass accretion. \textbf{Bottom right}: evolution of the initially less massive star from the time of the first BH formation until the second BH formation.  In all panels, the top black line indicates the stellar mass, the nuclear energy generation rate is colour-coded, while different hatching patterns correspond to different mixing processes: convection (green), overshooting (purple), red (semiconvection), thermohaline (yellow). The pink and red dotted lines indicate the mass of the helium and carbon core, respectively.} 
    \label{example_model} 
\end{figure*}

\begin{figure*}[!htbp]
    \centering
     \includegraphics[width=\linewidth]{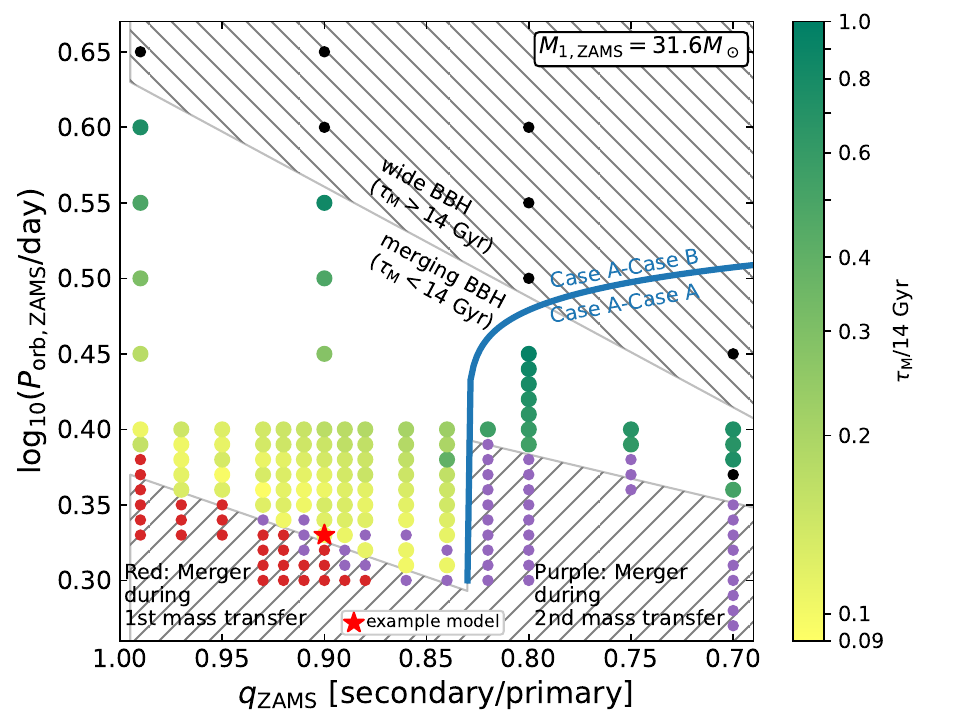}
    \caption{\textbf{Initial binary parameter space leading to merging BBH.} In the depicted initial mass ratio-initial orbital period ($q_{\rm ZAMS}$-log$_{\rm 10}\,P_{\rm orb,ZAMS}$) plane, each dot represents one detailed binary evolution models, with the initially more massive star starting with 31.6$\mso$. 
    Red and purple dots correspond to mergers during the first and second mass transfer respectively (bottom hatching). Black dots represent BBHs with merger timescales ($\tau_{\rm M}$) above 14\,Gyr (top hatching). The colourbar indicates the merger timescale in units of $14\,$Gyr for the obtained merging BBHs (unhatched background). Our example model is marked by a red star. The blue curve approximately separates Case A-Case A and Case A-Case B systems.
    }
    \label{grid_outcomes}
\end{figure*}

\begin{figure}
    \centering
    \includegraphics[width=\linewidth]{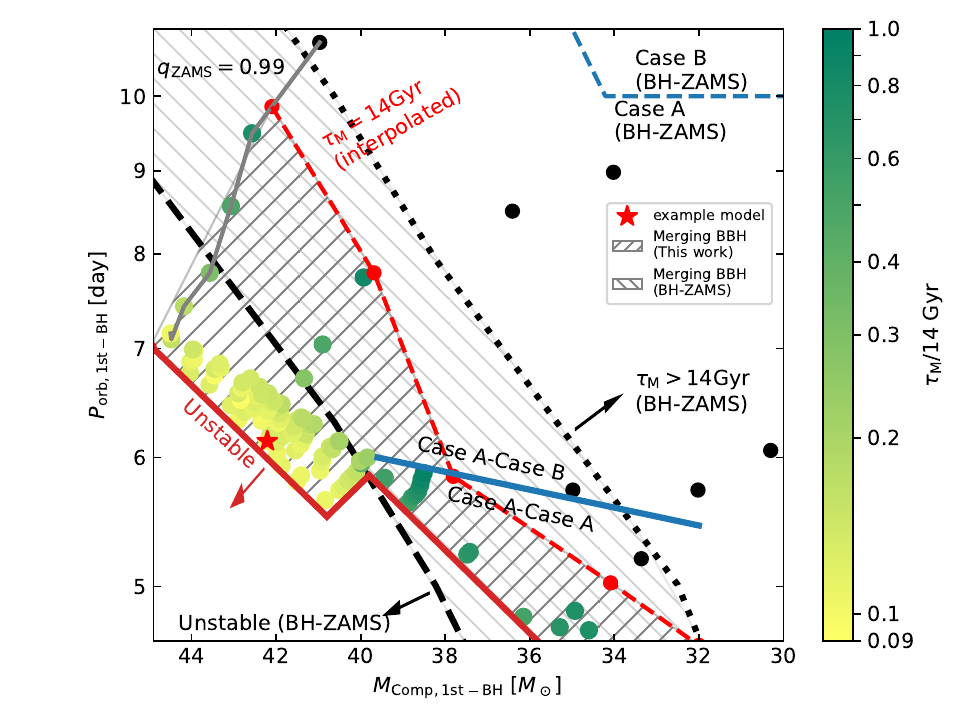}
    \caption{\textbf{Predicted BH-MS binaries and comparison with BH-ZAMS models.} Location of the models from our grid in the companion mass-orbital period plane at the time of the formation of the first BH (filled circles, where the colour reflects the merger time, as in Fig.\,\ref{grid_outcomes}).  The hatching patterns represent the parameter space of merging-BBH progenitors of our models (left-leaning diagonal and red borderlines), and of the BH-ZAMS models from ref\cite{Klencki2025arXiv250508860K} (right-leaning diagonal and black borderlines) computed with a BH mass of $13\mso$ (the first-born BHs in our models have masses 13--14$\mso$; cf., Supplementary Section \ref{app:BH-mass}).
    The solid grey line connects our models with $\qi=0.99$, which limits the allowed parameter space.
    The solid blue line separates models in our grid with Case\,A and with Case\,B mass transfer towards the BH companion. The blue dashed line is the corresponding line for the BH-ZAMS models. Our example model is plotted as a red star. }
     \label{BH-MS}
\end{figure}

\begin{figure}
    \centering
    \includegraphics[width=\linewidth]{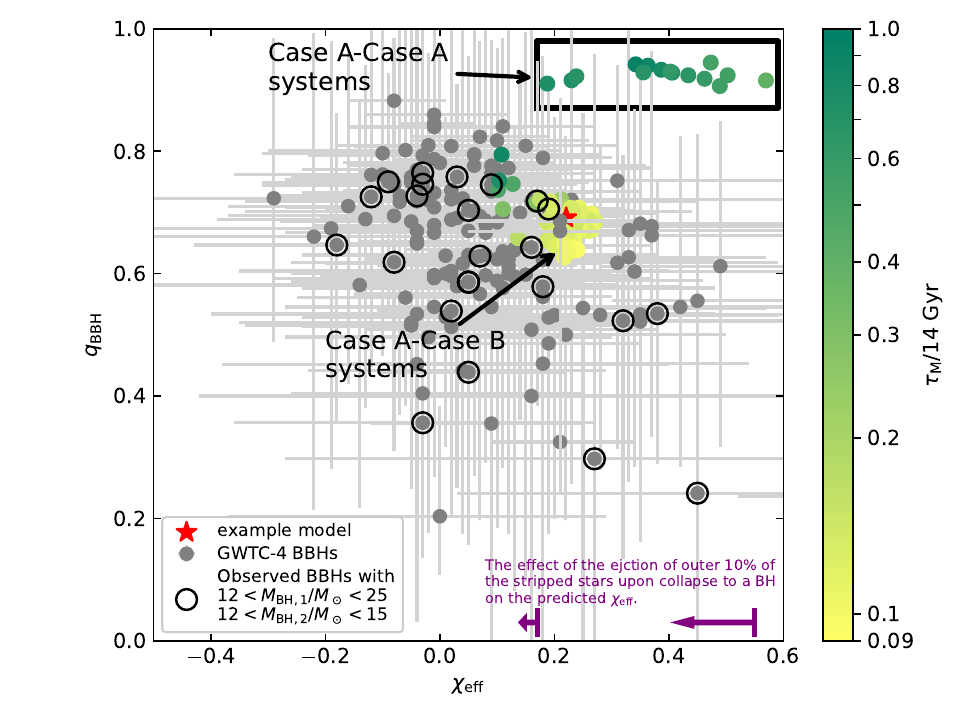}
    \caption{\textbf{Effective spins $\chi_{\rm eff}$ and mass ratios $q_{\rm BBH}$ of merging binary black holes.} The BBHs predicted by our models are indicated by coloured dots, with the colour representing the merger time as in Fig.\,\ref{grid_outcomes}. All our Case\,A-Case\,A models fall into the top right rectangle. We plot the effective spins and mass ratios obtained for the observed merging BBHs from the Gravitational-Wave Transient Catalogue 4 (GWTC-4; \cite{LVK2025arXiv250818082T}) 
    as gray dots with error bars, where black circles mark the observed BBHs with BH masses close to those of our merging BBH models (see legend). The red star corresponds to our example model. Purple arrows at the bottom indicate the effect of the ejection of the outer 10\% of the mass of the stripped stars upon collapse to a black hole on the predicted $\chi_{\rm eff}$ (Supplementary Section \ref{app:Xeff-mass-ejection}). 
    }
    \label{BH-BH}
\end{figure}

\section*{Methods}

\subsection*{Stellar evolution models}

Our stellar and binary evolution models are computed with the open-sources one-dimensional stellar evolution code MESA version 8845 (Modules for Experiments in Stellar Astrophysics; \cite{Paxton2011,Paxton2013,Paxton2015}). We adopt the same physical assumptions as those in refs.\cite{Brott2011}, \cite{Wang2020} and \cite{Xu2025arXiv250323876X}. We refer to these references for detailed description. The assumptions and input physics most relevant to this study are outlined as follows.

The merger timescale of merging BBHs suggests that their progenitors are likely born in low-metallicity environments \cite{Belczynski2016Natur.534..512B}. We accordingly compute our models with a metallicity tailored for the Small Magellanic Cloud, which is about 1/5th of the solar metallicity. This value roughly corresponds to the average metallicity of a redshift about 3--5 \cite{Langer2006,Kewley2007}, and is therefore appropriate for exploring the formation of merging BBHs. We adopt the metal distribution determined in ref.\cite{Brott2011}.

The size of convection zone is determined by the Ledoux criterion, which accounts for the effect of chemical gradient, in contrast to the Schwarzschild criterion. The convective mixing is modelled by the standard mixing-length theory with a parameter of $\alpha_{\rm MLT}=1.5$. Semiconvection takes place in the layer that is stable against convection according to the Ledoux criterion but unstable according to the Schwarzschild criterion, which is modelled following the method in ref.\cite{Langer1991} with $\alpha_{\rm SC}=1$ \cite{Langer1991,Brott2011}. The convective core of massive stars is expected to be larger than the expectation of the Ledoux criterion, as material overshoots the boundary of convective zone due to inertia. This convective overshooting is considered with a parameter of $\alpha_{\rm OV}=0.335$ as calibrated by ref.\cite{Brott2011}. We also take into account rotational mixing as those in ref.\cite{Heger2000}, containing the Eddington-Sweet circulation\cite{Eddington1926}, Goldreich-Schubert-Fricke instability\cite{Goldreich1967,Fricke1968},  and dynamical \cite{Endal1978} and secular shear instability \cite{Maeder1997}. The ratio of the turbulent viscosity to the diffusion coefficient $f_{\rm C}$ is taken to be $1/30$ \cite{Heger2000}. In addition, we take into account the effect of the internal magnetic field generated by rotation (Spruit-Taylor dynamo; \cite{Spruit2002}), which transports the rotational angular momentum from the core to the  envelope. 

We adopt the same stellar wind prescription as ref.\cite{Brott2011}. For hydrogen-rich stars with surface hydrogen mass fraction $X_{\rm H}$ above 0.7, we adopt the mass-loss rate prescription of refs.\cite{Vink2000,Vink2001}, which takes into account the bi-stability jump at about 22000\,K \cite{Vink1999}. We switch to the Wolf-Rayet mass-loss rate prescription of ref.\cite{Hamann1995} if $X_{\rm H}$ drops below 0.4. For $0.4\leq X_{\rm H}\leq0.7$, we apply a linear interpolation between these two regimes. The angular momentum loss through wind is calculated by removing the angular momentum contained in the layer stripped by the stellar wind (cf., Supplementary Section\,\ref{app:Xeff-wind}).  
For rotating massive stars, the surface gravity is reduced by  centrifugal force, enhancing the wind mass loss. We include this rotation-boosted mass loss rate as \cite{Friend1986ApJ...311..701F,Langer1997ASPC..120...83L}
\begin{equation}
    \frac{\dot{M}_{\rm w}}{\dot{M}_{\rm w}(\upsilon_{\rm rot}=0)} = \left[1-\left(\frac{\upsilon_{\rm rot}}{\upsilon_{\rm crit}}\right)\right]^{-0.43},
    \label{eq:rotation-boosted-wind}
\end{equation}
where $\dot{M}_{\rm w}$ is the wind mass loss rate, $\upsilon_{\rm rot}$ is the rotational velocity, and $\upsilon_{\rm crit}$ is the critical rotational velocity, which is evaluated as
\begin{equation}
    \upsilon_{\rm crit}^2 = \frac{GM}{R}(1-\Gamma),
\end{equation}
where $G$ is the gravitational constant, $M$ is the stellar mass, $R$ is the stellar radius, and $\Gamma$ is the Eddington factor. According to these formulae, one can expect a strong rotation-boosted wind when the radiation power approaches the Eddington limit $(\Gamma \rightarrow 1)$.

\subsection*{Binary evolution models}

We compute the evolution of two stellar components simultaneously. Following ref.\cite{Wang2020}, we assume both stars are tidally locked at zero-age main sequence, which is appropriate for close binaries. Tidal interaction is modelled according to ref.\cite{Hut1981}, allowing the exchange of angular momentum between stellar rotations and the orbit.  
The size of Roche lobe is determined by the volume-equivalent radius \cite{Eggleton1983}, and mass transfer occurs when the star fills its Roche lobe. We compute the mass transfer rate implicitly such that the mass donor is limited within its Roche lobe.

During the first mass transfer phase, we expect binary mergers if stellar components overflow the outer Lagrangian point, if the mass transfer towards a stripped star occurs, and if mass transfer rate exceeds an artificial upper limit 0.1$\msoy$. The transferred material carries angular momentum, and spins up the main-sequence mass gainer. Once the mass gainer reaches critical rotation, further accretion is prevented by increasing wind mass-loss rate such that the star remains below critical rotation, and the accretion efficiency is regulated by the spin-up through accretion and  spin-down by tidal torque and wind mass loss.
The non-accreted material is assumed to be ejected in the form of isotropic winds launched by the radiation power of the system. We also expect a merger if the radiation of the system is not strong enough to eject the non-accreted material \cite{Wang2020,Langer2020,Henneco2024,Xu2025arXiv250323876X}. 

Once the initial primary star produces a BH (i.e., reaching core helium depletion), we save the model of the secondary star at the same evolutionary moment, which serves as the initial model for the second mass transfer phase. This treatment retains the structuring changing caused by the mass accretion during the first mass transfer phase. The corresponding orbital period and BH mass are set according to our assumption on BH formation (see below). The tidal torque acting on the BH is ignored. Different from a main-sequence mass gainer, we assume BH accretion is limited by the Eddington accretion rate, which is calculated by the formula in ref.\cite{Podsiadlowski2003MNRAS.341..385P}. Due to its compactness, a BH can efficiently convert the gravitational potential energy of the transferred material into radiation \cite{Frank2002apa..book.....F}, and we accordingly assume that a BH mass gainer is always capable of expelling the non-accreted material out of the binary system. The spin evolution of the BH is computed by assuming that the BH accretes the specific orbital angular momentum at its innermost stable circular orbit, whose radius $R_{\rm ISCO}$ is given by 
\begin{equation}
    R_{\rm ISCO} = \frac{6GM_{\rm BH}}{c^2},
\end{equation}
where $M_{\rm BH}$ is the BH mass, and $c$ is the speed of light. The specific orbital angular momentum, $j_{\rm ISCO}$, is assumed to be Keplerian, 
\begin{equation}
    j_{\rm ISCO}=\sqrt{GM_{\rm BH}R_{\rm ISCO}}.
\end{equation}
General relativistic corrections are ignored, as they do not considerably affect our conclusions. 
We do not include the accretion through wind capture (see also \cite{Sen2021,Sen2024A&A...690A.256S}).
In addition, we expect the merger of a star and a BH if the mass transfer rate is higher than our artificial upper limit $10\msoy$.

\subsection*{Black hole formation}

For simplicity, in this study we adopt a different BH formation prescription from ref.\cite{Xu2025arXiv250323876X}. Our stellar models are terminated at core helium depletion. After this point, stars evolve very rapidly, and we do not expect significant changes in stellar and orbital properties at the SMC metallicity. Hence 
core helium depletion is sufficiently close to core collapse for the purpose of this study. 
Following ref.\cite{Langer2020}, we assume the entire star at core helium depletion collapses into a BH without mass ejection or a natal kick. Consequently, the orbital period remains unchanged across BH formation. In addition, we assume that no rotational angular momentum is lost during the collapse  (but see Supplementary Section \ref{app:Xeff-mass-ejection}). Therefore, the dimensionless spin parameter $a_{\rm spin}$ of the BH is equal to that of the progenitor star at core helium depletion, which is given by \cite{Marchant2024A&A...691A.339M} 
\begin{equation}
    a_{\rm spin} = \text{min}\,\left[1,\,\frac{Jc}{GM^2}\right],
\end{equation}
where $J$ is the total angular momentum of the pre-collapse star.

\subsection*{Binary black holes}

As a consequence of our BH formation prescription, the birth orbital period of a BBH is equal to the final orbital period of its progenitor system. The BH spins are expected to be aligned with the orbital angular momentum (but see  \cite{Tauris2022,Baibhav2024arXiv241203461B,LARSEN2025102459}). Then the effective spin parameter of the merging BBH $\chi_{\rm eff}$ is given by
\begin{equation}
    \chi_{\rm eff} = \frac{M_{\rm BH,1st}\,a_{\rm spin,1st}+M_{\rm BH,2nd}\, a_{\rm apin,2nd}}{M_{\rm BH,1st} + M_{\rm BH,2nd}},
    \label{Xeff}
\end{equation}
where $M_{\rm BH,1st}$ and $M_{\rm BH,2nd}$ are the masses of the first- and second-born BHs, $a_{\rm spin,1st}$ and $a_{\rm spin,2nd}$ are their dimensionless spin parameters respectively. 
It has been shown that BH spins are not considerably affected by tidal interaction during the late spiral-in phase of BBH mergers\cite{Campanelli2006PhRvD..74h4023C}. We accordingly assume that the effective spin parameter remains fixed throughout the orbital evolution of the BBHs, which is solely driven by gravitational wave radiation \cite{PetersPhysRev.136.B1224}.  We calculate the merger timescale, which is the time from the formation of a BBH until its final merger, by using the formula in ref.\cite{Mandel2021RNAAS...5..223M}, which is an accurate analytical fitting formula to the solution derived by ref.\cite{PetersPhysRev.136.B1224}.
In order to compare with the mass ratios of observed merging BBHs, we calculate the mass ratio of the predicted BBHs $q_{\rm BBH}$ as 
\begin{equation}
    q_{\rm BBH}=\text{min}\left[\frac{M_{\rm BH,1st}}{M_{\rm BH,2nd}},\,\frac{M_{\rm BH,2nd}}{M_{\rm BH,1st}}\right],
\end{equation}
ensuring that $q_{\rm BBH}$ is always below unity.

\section*{Code and Data Availability}

The MESA code is publicly available at \url{https://docs.mesastar.org/}.
The model data is available on request. 
The necessary MESA files for reproducing our models will be available upon acceptance of the paper.

\section*{Acknowledgments}

We thank Linhao Ma for helpful discussion.
X.-T.X. thanks Wenjie Wu for the help with the transfer of model data.
This research was supported in part by grant NSF PHY-2309135 to the Kavli Institute for Theoretical Physics (KITP).
X.-T.X. is supported by the Tsung-Dao Lee postdoctoral fellowship at the Tsung-Dao Lee Institute (TDLI).
X.-D.L. and C.W. are supported by the National Key Research and Development Program of China (2021YFA0718500), the Natural Science Foundation of China under grant No. 12041301 and 12121003.

\section*{Author contributions}

X.-T.X. constructed the full-evolution model based on the work by C.W., computed the model grid, wrote most of the manuscript jointly with N.L.. J.K. provided the predictions obtained with the BH-ZAMS models. All authors contributed to the analysis and interpretation of the results, and the writing of the manuscript.

\section*{Correspondence}
Correspondence and requests for materials should be addressed to Xiao-Tian Xu\\ (email: xxu-tdli@sjtu.edu.cn; xxu@astro.uni-bonn.de).

\section*{Competing Interests}
The authors declare no competing interests.

\bibliography{Xu_SMC}
\bibliographystyle{naturemag}
\setcounter{section}{0}

\section*{Supplementary Information}
\renewcommand{\thesection}{\Alph{section}}
\section{Uncertainties in black hole formation\label{app:BH-formation}}

In our model grid, we have fixed the initial mass of the primary star to $31.6\mso$ (1.5 in log), which is commonly adopted in past detailed simulations on merging binary black holes (BBHs) \cite{Marchant2021,Gallegos-Garcia2021,Klencki2025arXiv250508860K}. We expect this primary star to form a BH, according to previous detailed simulations on the explodability of massive stars \cite{Fryer1999ApJ...522..413F,Woosley2002RvMP...74.1015W,Heger2003ApJ...591..288H,Sukhbold2018}. Our $31.6\mso$ primary star produces an helium core of $13\text{--}14\,\mso$, corresponding to a compactness parameter of about 0.6 according to the high-resolution models in ref.\cite{Sukhbold2018ApJ...860...93S}, 
which is sufficiently high for direct collapse. For the secondary stars in our models, they all accrete mass above $30\mso$ during the first mass transfer phase, implying that all systems shown in the main text Fig.\,\ref{BH-MS} are expected to be BBHs progenitors. 

\begin{figure}[!ht]
    \centering
    \includegraphics[width=0.8\linewidth]{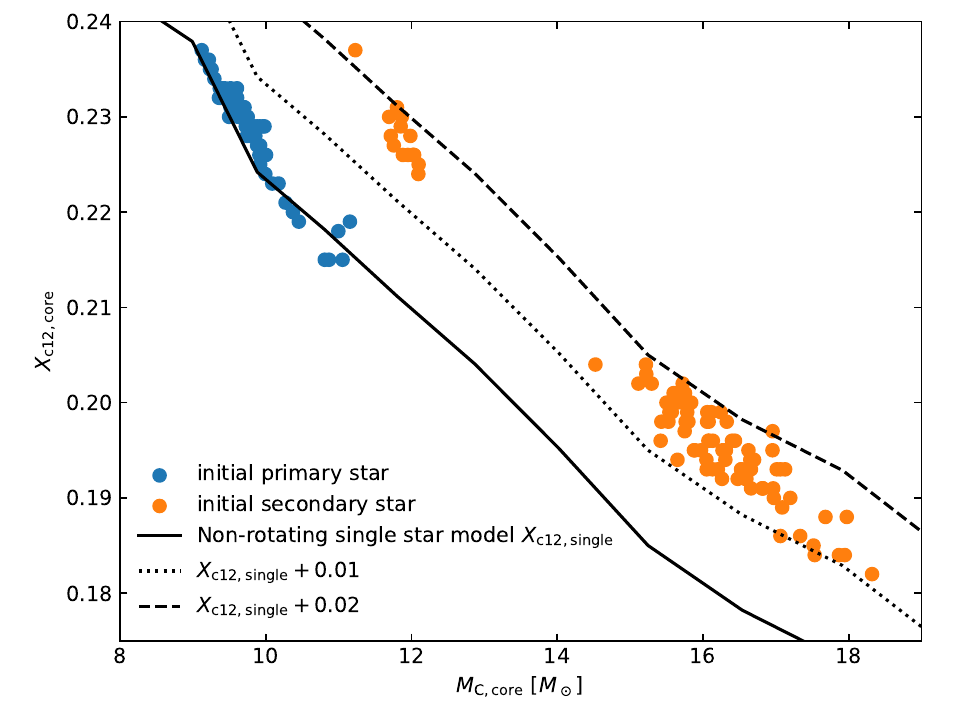}
    \caption{\textbf{Core carbon mass fraction $X_{\rm c12,core}$ and carbon core mass $M_{\rm C,core}$  at core helium depletion of the progenitors of the merging BBHs presented in the main text Fig.\,\ref{BH-BH}.} The blue and orange dots correspond to initial primary and secondary stars respectively. The black solid line $X_{\rm c12,single}$ is the result of the non-rotating single star models in ref.\cite{Xu2025arXiv250323876X} computed with the same input physics as our full-evolution models. The black dotted and dashed lines are $X_{\rm c12,single}+0.01$ and $X_{\rm c12,single}+0.02$, plotted for comparison purpose.}
    \label{Mco-C12}
\end{figure}

Binary interaction has been proposed to have critical effects on the explodability of massive stars \cite{Schneider2021,Schneider2023,Schneider2024A&A...686A..45S,Laplace2025A&A...695A..71L,Maltsev2025A&A...700A..20M}, may account for the multi-modal feature of the observed chirp mass distribution \cite{Willcox2025arXiv250820787W,Willcox2025arXiv251007573W}. The binary stripping of the hydrogen-burning shell can alter the mixing process inside the helium core, 
leading to a different core carbon mass fraction at core helium depletion in contrast to single-star counterparts, which changes the outcome of carbon and neon burning, consequently altering the explodability \cite{Patton2020MNRAS.499.2803P,Laplace2025A&A...695A..71L}. 
To address this uncertainty, in Fig\,\ref{Mco-C12}, we present the core carbon mass fraction and the carbon core mass at core helium depletion of our merging BBH progenitors, and compare them to the non-rotating single star models in ref.\cite{Xu2025arXiv250323876X}, which are computed with the same input physics as our full-evolution models. Different from ref.\cite{Schneider2021}, we find that the core carbon mass fraction at core helium depletion of the initial primary star remains consistent with the single-star results. This is likely the consequence of the remaining hydrogen above the stripped star (see also \cite{Laplace2020}). By contrast, the hydrogen-burning shell of the initial secondary stars can be removed completely (the main text Fig.\,\ref{example_model} and Supplementary Fig.\,\ref{kipp:CaseA-CaseA}), resulting in a upward shift of 0.01--0.02 in core carbon mass fraction relative to the single-star models, which is much weaker than that reported in ref.\cite{Schneider2021}. We therefore do not expect binary stripping to largely affect the explodability of the star in our model grid.

Massive BHs are expected to be formed through direct collapse \cite{Fryer1999ApJ...522..413F}, and we accordingly assume the entire star at core helium depletion collapses into a BH, consistent with refs.\cite{Langer2020,Marchant2021,Klencki2025arXiv250508860K}. However, some mass ejection might occur during the formation of relative low-mass BHs \cite{Fryer1999ApJ...522..413F,Heger2003ApJ...591..288H}, which would slightly widen the orbit and induce eccentricity. Ref.\cite{Xu2025arXiv250323876X} shows that the eccentricity of BH-MS binaries caused by mass ejection is roughly below 0.1 by assuming 20\% of the mass of the helium-rich envelope is ejected during BH formation. Therefore, if mass ejection were included, the merger timescales would be slightly longer than our fiducial model.

\section{The choice of initial parameters \label{app:Pq-window}}

\subsection{ZAMS orbital periods\label{app:Porb-window}}

In this study, we restrict our analysis to close binaries, as we do not expect wide black hole-main-sequence star (BH-MS) binaries to form merging binary black holes through stable mass transfer. It is possible if the mass donor is much more massive than the BH according to previous BH-ZAMS models \cite{Marchant2021,Gallegos-Garcia2021,Klencki2025arXiv250508860K}. However, the most extreme mass ratio of wide BH-MS binaries expected by our full-evolution model is about 0.5 (BH/MS), achieved near $\qi=1$. The reason is that a main-sequence mass gainer in initially wide binaries can hardly gain mass due to the rotation-limited accretion efficiency \cite{Packet1981,Langer2020,Rocha2024ApJ...971..133R,Xu2025arXiv250323876X}, and the mass donor loses about half of its initial mass \cite{Langer2020,Xu2025arXiv250323876X}. This mass ratio is not extreme enough to shrink wide binaries below the Hubble time. However, we note that possibility is not excluded theoretically due to the uncertainties in accretion efficiency in wide binaries \cite{Petrovic2005,Shao2014,Rocha2024ApJ...971..133R,Schurmann2025arXiv250323878S,Xu2025arXiv250323876X,Lechien2025arXiv250514780L,Bao2025arXiv250602662B,Zapartas2025arXiv250812677Z} and supernova physics\cite{Fryer1999ApJ...522..413F,Fryer2012ApJ...749...91F,Schneider2019,Willcox2025arXiv251007573W}, which lie beyond the scope of this study.

In addition, the population synthesis calculation at solar metallicity performed with the POSYDON framework suggests that merging BBHs may also be formed through stable mass transfer by initially wide binaries even with a ZAMS mass ratio of 0.4 \cite{Bavera2023NatAs...7.1090B}. Since the POSYDON models also assume a rotation-limited accretion efficiency, the reason for the different parameter space likely lies in their BH formation prescription, allowing extreme mass ratios for BH-MS binaries, and assumption on tides that, when the mass donor in an eccentric binary fills the Roche lobe at periastron, the orbit is assumed to be instantly circularised to an new circular orbit with the orbital separation equal to the separation at periastron\cite{Bavera2023NatAs...7.1090B}.

\subsection{ZAMS mass ratios\label{app:q-window}}

We have not considered the models with $\qi$ below 0.7 because  most such low-$\qi$ close binaries are expected to merge during the first mass transfer phase according to past detailed binary evolution models \cite{Wellstein2001,Wang2020,Langer2020,Fragos2023ApJS..264...45F,Henneco2024,Xu2025arXiv250323876X}.  
The upper limit of $\qi$ of our model grid is taken to be 0.99. 
While our equal-mass binaries merge due to the overflow through the outer Lagrangian point during contact phase, the models with $\qi=0.99$ avoid merging. We do not resolve the transition between surviving systems and mergers, as the parameter space from 0.99 to 1 is small.

\subsection{ZAMS stellar rotational velocities\label{app:rotation}}

The initial rotational velocities of the stellar components in our models are set by tidal locking at zero-age. As we show here, this choice of initial rotational velocities does not considerably affect the parameter space of merging BBHs. 

In close binaries, tidal interaction causes an exchange between stellar rotational angular momentum and  orbital angular momentum. For an initially fast-rotating star, rotational angular momentum is transferred to the orbit, causing additional orbital widening. Conversely, an initially slow-rotating star leads to additional orbital shrinkage. In our Case A-Case B example model (Fig.\,\ref{example_model}), the initial orbital angular momentum is about $4\times10^{54}\,\text{g}\,\text{cm}^2\,\text{s}^{-1}$, while the initial rotational angular momentum of the stellar components is about $3\text{--}5\times10^{52}\,\text{g}\,\text{cm}^2\,\text{s}^{-1}$, which is about two orders of magnitude lower the orbital angular momentum. Therefore, we do not expect noticeable shift in the parameter space when adopting different initial rotational velocities.

Initially fast-rotating stars may enter chemically homogenous evolution (CHE) due to rotational mixing \cite{Heger2000,Brott2011}. 
The parameter space of the CHE channel towards merging BBHs appears to be very limited according to previous detailed simulations \cite{Marchant2016,Hastings2020A&A...641A..86H,Popa2025arXiv250900154P}. It typically requires initial stellar mass above 70$\mso$, low metallicity ($\leq1/10$th solar metallicity \cite{Marchant2016}), and very short initial orbital periods such that stars can already fill the Roche lobe at zero-age main sequence. Therefore we do not expect the binaries considered in this work to form merging BBHs through the CHE channel, when initial rotational velocities are varied in a reasonable range.

\section{Uncertainties in the predicted black hole spins\label{app:Xeff-uncertainties}}

\subsection{Mass ejection during collapse\label{app:Xeff-mass-ejection}}

In the main text, we have assumed that progenitor stars directly collapse into BHs without any mass ejection or angular momentum loss. While this assumption has only minor effects on the post-collapse orbital properties (Supplementary Section \ref{app:BH-formation}), it could overestimate the resulting BH spins, as the specific angular momentum of the pre-collapse star increases rapidly near the surface of the star (e.g., \cite{Yoon2005A&A...443..643Y}), and the ejection of the outer layers may considerably reduce the rotational angular momentum preserved in the BH.
Here we estimate the effects of mass ejection by assuming that the outer 10\% of the stripped stars is ejected during BH formation. Since the fraction of mass ejection is assumed to be the same for both first- and second-born BHs, the mass ratio of merging BBHs is unchanged, and, in order to focus on the changes in effective spins, we ignore the changes in orbital separation and merger timescales related to mass ejection. 

\begin{figure}[!ht]
    \centering
    \includegraphics[width=0.8\linewidth]{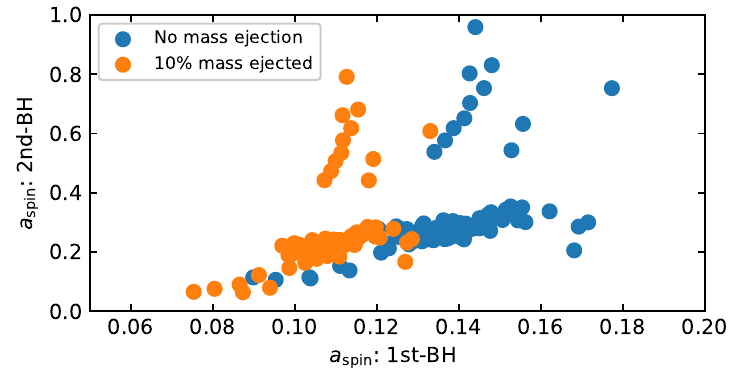}
    \includegraphics[width=0.8\linewidth]{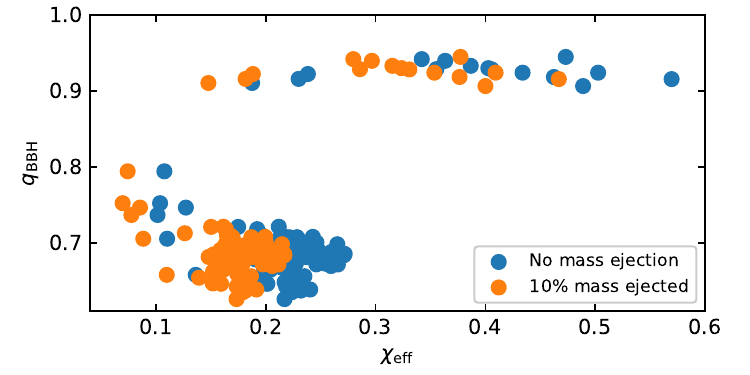}
    \caption{\textbf{Effects of mass ejection during BH formation on the spin parameters of the BHs.} In both panels, blue and orange dots correspond to the predictions without/with mass ejection during BH formation. \textbf{The upper panel} presents the predicted spin parameters, $a_{\rm apin}$, of the first- and second-born BHs. \textbf{The lower panel} presents the effective spin parameter, $\chi_{\rm eff}$, and mass ratios, $q_{\rm BBH}$, of merging BBHs.}
    \label{massejecion-aspin}
\end{figure}

Figure\,\ref{massejecion-aspin} shows that mass ejection can reduce the spin parameters by $\sim$20--30\% compared to the direct collapse case, and the effective spin parameter of merging BBHs is reduced by a similar factor.  Hence, the shape of the $\chi_{\rm eff}\text{--}q_{\rm BBH}$ relation of Case A-Case A and -Case B BBHs remain unchanged. Generally, the effects of mass ejection are more significant for BHs with higher spins. 

\subsection{Wind angular momentum loss scheme\label{app:Xeff-wind}}

The angular momentum preserved in the pre-collapse star is affected by the angular momentum loss related to the wind during the core helium burning phase, whose mass loss rate is still poorly constrained \cite{Grafener2017}. Meanwhile, the specific angular momentum carried by wind material also remains uncertain \cite{Nathaniel2025arXiv250212107N,Xu2025arXiv250323876X}. Our models adopt the same numerical treatment for wind angular momentum loss as in refs.\cite{Chen2022ApJ...930..134C,Xu2025arXiv250323876X} that the angular momentum lost in one time step is calculated by removing the angular momentum contained in the layer stripped by the stellar wind (the "old" scheme), which is the default setting in MESA version 8845. In later versions, the angular momentum loss is calculated by multiplying the mass loss rate by the surface specific angular momentum (the "new" scheme), resulting in a stronger spin-down torque. The new scheme yields an angular momentum loss independent of time step choices, which appears more physical. However, the old scheme better reproduces the observed rotational velocities of Galactic O stars \cite{Nathaniel2025arXiv250212107N}, suggesting that the new scheme likely overestimates the angular momentum in the wind-removed layers (see Fig. 1 in ref.\cite{Nathaniel2025arXiv250212107N}).

To assess this uncertainty, we rerun our Case A-Case B example model (Fig.\,\ref{example_model} and Supplementary Section \ref{app:kipp-CaseA-CaseB}) and Case A-Case A example model (Supplementary Section \ref{app:CaseA-CaseA}) with the new scheme. The resulting spin parameters are presented in Tab. \ref{tab:AML-scheme}, and the spins computed by the old scheme are listed for comparison. For the Case A-Case B model, the spins of both BHs are reduced by about 30\% comparing to the old scheme, which is comparable to the effects of mass ejection (Supplementary Section \ref{app:Xeff-mass-ejection}). However, for the Case A-Case A model, the spin of the first-born BH is reduced by 20\%, while that of the second-born BH reduced by about 50\%. With the new scheme, the Case A-Case A model still has a higher effective spin parameter than that of the Case A-Case B model. Hence, we would still expect distinct  $\chi_{\rm eff}\text{--}q_{\rm BBH}$ relations for Case A-Case A and -Case B merging BBHs if the new scheme was adopted.

\begin{table}[!ht]
    \centering
    \caption{\textbf{Effects of angular momentum loss scheme on black-hole spin parameters.}
    \label{tab:AML-scheme}}
    \begin{tabular}{c|ccc|ccc}\hline\hline
        Type&\multicolumn{3}{l|}{Case A-Case B} &\multicolumn{3}{l}{Case A-Case A}\\
       $(\mi,\qi,\porbi)$ & \multicolumn{3}{l|}{$(31.6\mso,\,0.9,\,2.1\,\text{d})$} & \multicolumn{3}{l}{$(31.6\mso,\,0.82,\,2.5\,\text{d})$}\\\hline
       &$a_{\rm spin,1}$&$a_{\rm spin,2}$&$\chi_{\rm eff}$&$a_{\rm spin,1}$&$a_{\rm spin,2}$&$\chi_{\rm eff}$\\\hline
       Old scheme &0.17&0.27& 0.22&0.14&0.80&0.49\\
       New scheme &0.12&0.19&0.16&0.11&0.42&0.28\\
       \hline
    \end{tabular}
    \begin{flushleft}
    In this table, $\mi$ is the ZAMS primary mass, $\qi$ is the ZAMS mass ratio, $\porbi$ is the ZAMS orbital period, $a_{\rm spin,1}$ is the spin parameter of the first-born BH, $a_{\rm spin,2}$ is the spin parameter of the second-born BH, and $\chi_{\rm eff}$ is the effective spin parameter of the merging BBH.   
    \end{flushleft}
\end{table}

\section{Further details of our model grid\label{app:grid-details}}

\subsection{\texorpdfstring{Structure of our rejuvenated $42\mso$ model}{Structure of our rejuvenated 42Msun model}} \label{app:kipp-single-and-rejuvenated}

\begin{figure}[!ht]
    \centering
    \includegraphics[width=0.9\linewidth]{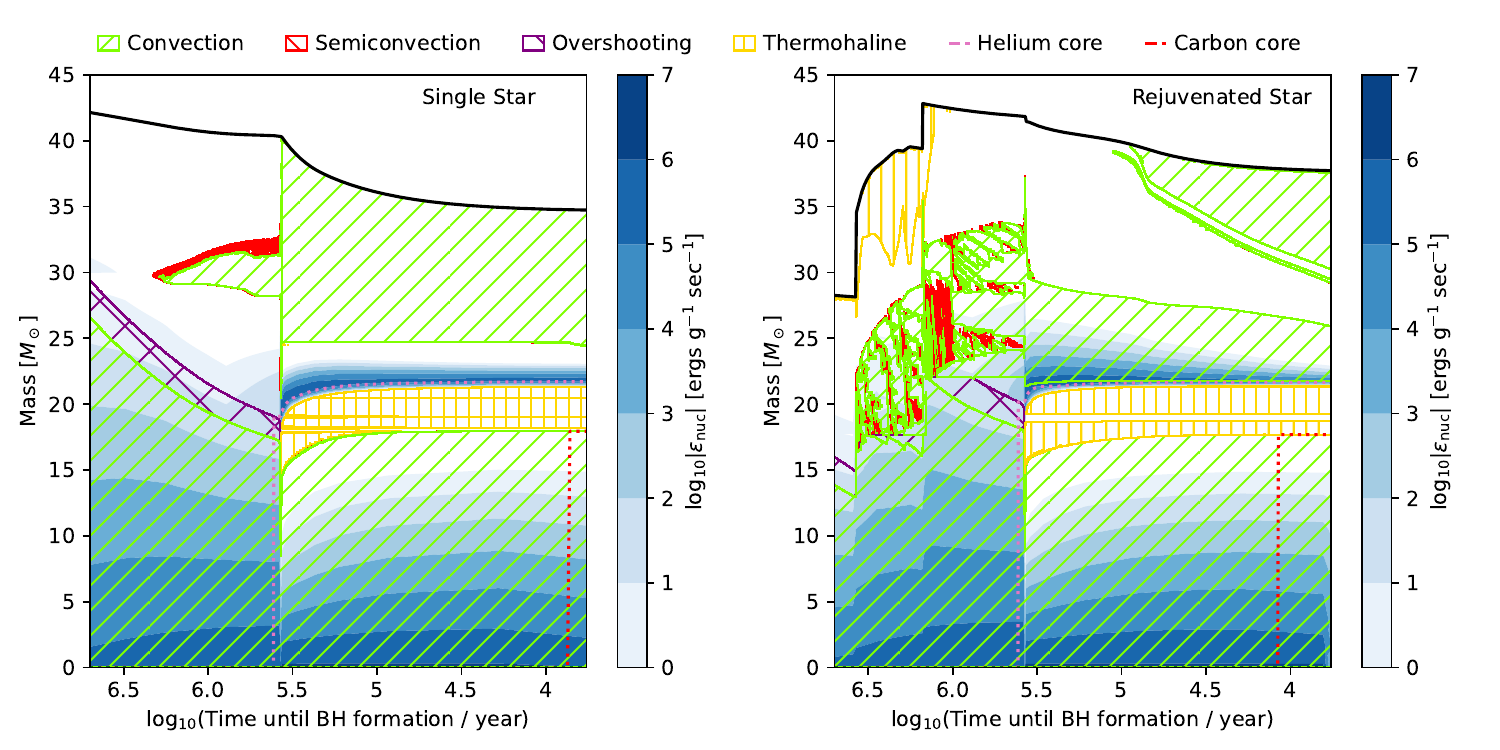}
    \caption{\textbf{Kippenhahn diagrams of a 42$\mso$ single star (left) and a 42$\mso$ rejuvenated star evolving in isolation after the formation of the first-born black hole (right).} The meaning of lines, colours, and hatching patterns are  the same as Supplementary Fig.\,\ref{kipp:example_model}}
    \label{kipp:rejuvenated_star_isolation_and_normal_single_star}
\end{figure}

To better illustrate the difference in stellar structure between rejuvenated stars and single stars, we compute a binary evolution model with ZAMS parameters of $M_{\rm 1,ZAMS}=31.6\mso$, $q_{\rm ZAMS}=0.9$, and $P_{\rm orb,ZAMS}=2.1\,\text{d}$. 
Unlike our full-evolution models, we evolve the $42\mso$ rejuvenated star in isolation after the formation of the first-born BH.
For comparison, we also evolve a non-rotating $42\mso$ single star with the same input physics. We presents the Kippenhahn diagrams in Fig.\,\ref{kipp:rejuvenated_star_isolation_and_normal_single_star} (left: single star; right: rejuvenated star).

A single $42\mso$ MS star has a convective core about 20--25$\mso$ with a radiative envelope. Near core hydrogen depletion, a thin convective region, containing 2--3$\mso$ mass, appears at the middle of its envelope. The rejuvenated star, in contrast, develops a convective zone with about 10$\mso$ above its core due to mass accretion, consistent with previous simulations \cite{Langer1991,Renzo2021ApJ...923..277R,Miszuda2025}, and accreted material induces thermohaline mixing near the surface.
After core hydrogen depletion, the single star develops a strongly convective envelope, whereas the envelope of the rejuvenated star remains radiative until about $10^{5}$\,yr before the BH formation. 

\begin{figure}[!ht]
    \centering
    \includegraphics[width=0.45\linewidth]{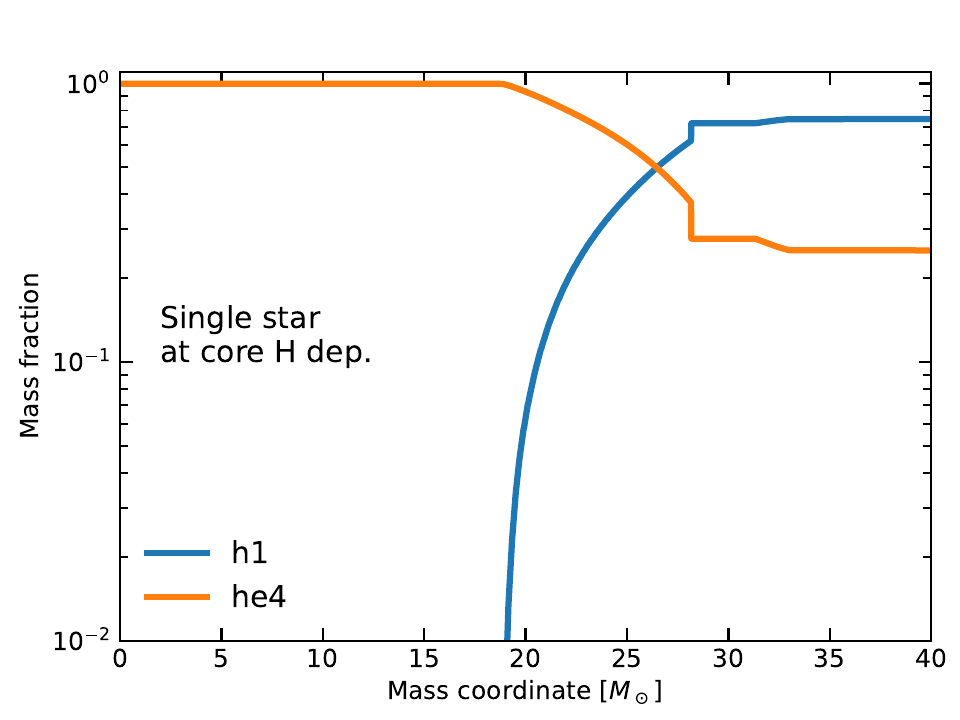}
    \includegraphics[width=0.45\linewidth]{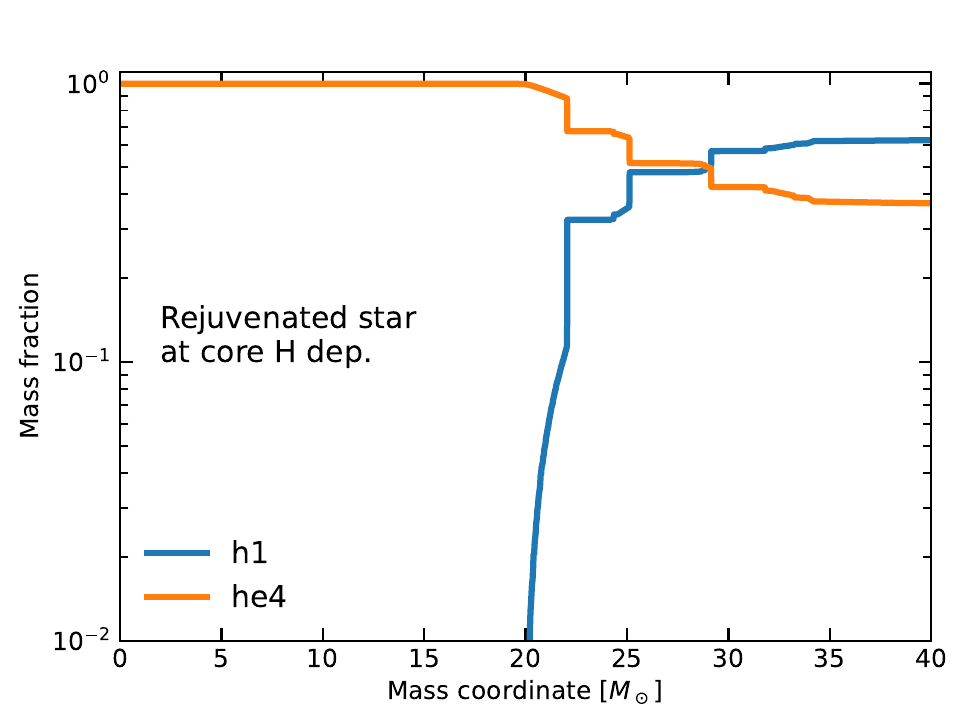}
    \caption{\textbf{Chemical abundance at core hydrogen depletion as a function of mass coordinate}. The left and right panels correspond to the single star and rejuvenated star models presented in Supplementary Fig.\,\ref{kipp:rejuvenated_star_isolation_and_normal_single_star}. The X-axis is the mass coordinate, and the Y-axis is the mass fraction. Colours correspond to different isotopes (blue for hydrogen; orange for helium). }
    \label{chemical_strucutre_at_H_dep}
\end{figure}

Figure\,\ref{chemical_strucutre_at_H_dep} compares chemical structures at core hydrogen depletion. Below about $18\mso$, both stars' hydrogen-depleted cores are similar. Above the helium core, hydrogen mass fraction $X$ rises, and helium mass fraction $Y$ declines. The off-centre convective zones result in several chemically homogenous regions in the envelope of the rejuvenated star. At surface, the rejuvenated star reaches $Y=0.4$ but $Y=0.25$ for the single star, consistent with previous works \cite{Langer2020,Renzo2021ApJ...923..277R,Sen2022,Sen2023,Xu2025arXiv250323876X,Jin2025}.

\subsection{Further details of the Case A-Case B example model\label{app:kipp-CaseA-CaseB}}

\begin{figure*}[!ht]
    \centering
       \includegraphics[width=\linewidth]{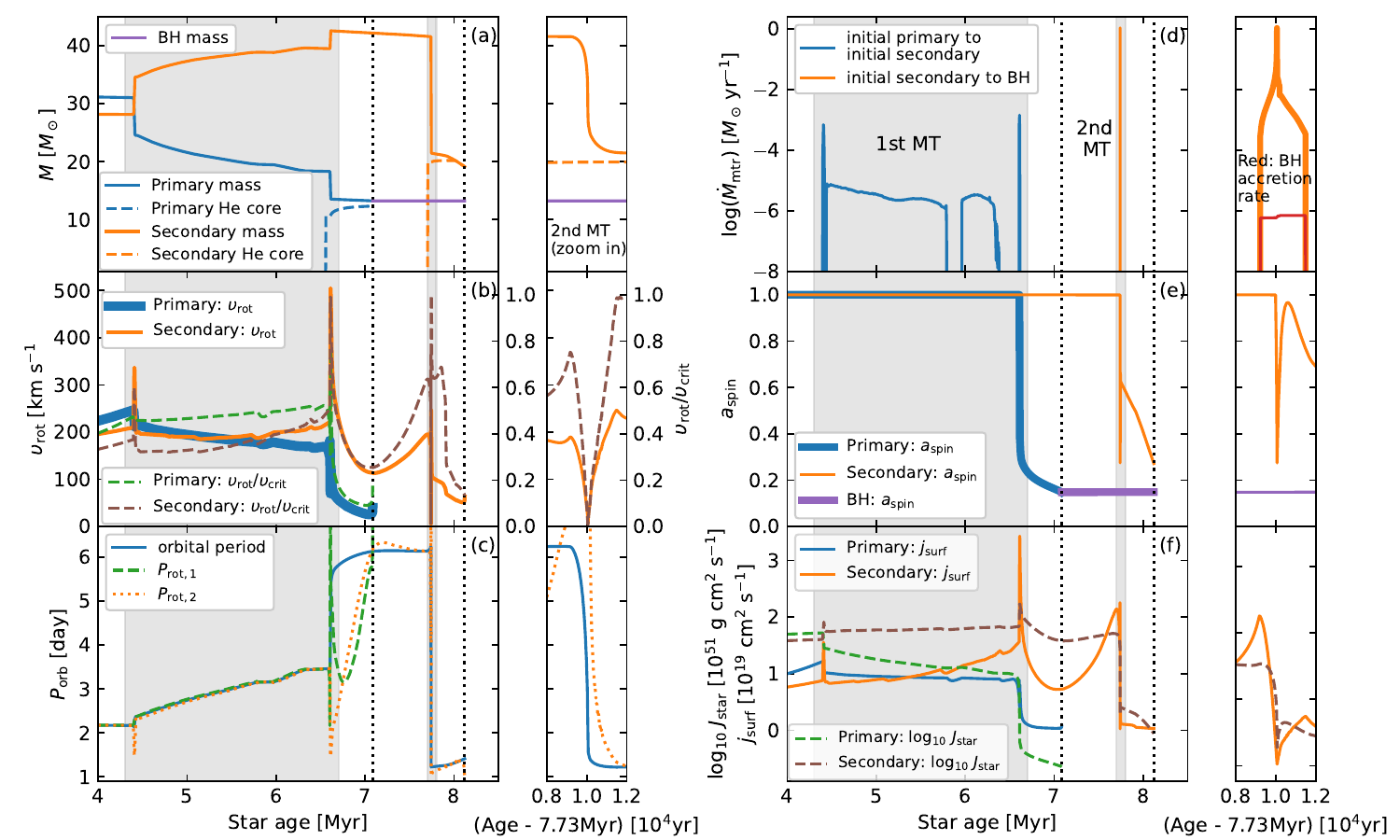}
    \caption{\textbf{Evolution of the Case A-Case B  example model.} This model has an initial primary mass of $31.6\mso$, an initial mass ratio of 0.9, and an initial orbital period of 2.1\,d. In all panels, the two vertical dashed lines mark the formation time of black holes. The narrow columns provide a zoom-in for the evolution during the second mass transfer. The grey background indicates two mass transfer phases, which are wider than the actual duration for presentation purpose.We use blue lines for properties of the initially more massive star, and orange lines for its companion.  \textbf{Panel (a)} shows the masses of stellar components (solid lines), and their helium core masses (dashed). The purple line represents the mass of the first-born black hole. \textbf{Panel (b)} shows the evolution of surface rotational velocities $\upsilon_{\rm rot}$  and the ratio of $\upsilon_{\rm rot}$ to the critical rotational velocity $\upsilon_{\rm crit}$ (green dashed line: primary  star; brown dashed line: secondary star). \textbf{Panel (c)} shows the evolution of orbital period and the rotational periods of stars (initial primary: green dashed line; initial secondary: orange dotted line). \textbf{Panel (d)} shows the evolution of the mass transfer rate, where the direction of mass transfer is distinguished by colours (blue: towards the initial secondary star; orange: towards the first-born BH). In addition, the mass accretion rate of the BH is plotted in red in the zoom-in panel. \textbf{Panel (e)} presents the evolution of dimensionless spin parameter $a_{\rm spin}$ (blue: primary star; orange: secondary; purple: first-born BH). \textbf{Panel (f)} shows the surface specific angular momentum $j_{\rm surf}$ in the unit of $10^{19}\text{\,cm}^{2}~\text{s}^{-1}$ and logarithmic total rotational angular momentum $J_{\rm star}$ in the unit of $10^{51}\,\text{g}\,\text{cm}^{2}\,\text{s}^{-1}$. }
    \label{kipp:example_model}
\end{figure*}

In this section, we provide a more detailed description of the model (Fig.\,\ref{kipp:example_model}). At the beginning of the calculation, both stars are assumed to be tidally locked at ZAMS, resulting in rotational velocities of about 200$\kms$ for both stars. The stars expand gradually until 4.4\,Myr, when the primary star fills its Roche lobe, triggering a fast Case A mass transfer phase. About $5\mso$ material is transferred to the secondary star, spinning up the secondary star to a rotational velocity of about 330$\kms$. During the subsequent nuclear-timescale Case A mass transfer, the orbital period increases from 2.1\,d to 3.5\,d, as the mass donor becomes less massive than the gainer. Envelope stripping during this phase slightly reduces the convective core of the mass donor. Meanwhile, the initial secondary star keeps gaining mass up to about $40\mso$. The transferred CNO-cycle processed material causes significant thermohaline mixing near the
surface of the secondary. As the accreted material mixes downwards, the star develops a thick convection zone above its core region due to local density enhancement \cite{Miszuda2025}. At 7.7\,Myr, fast expansion of the mass donor after core hydrogen depletion triggers another mass transfer phase (Case AB), which we still classify as the first mass transfer, since the initial primary star acts as the donor. It widens the orbit to 5.6\,d and spins up the mass gainer to its critical rotation. During this phase, 1.7$\mso$ material is ejected out of the binary system due to the limitation of critical rotation. Then the initial primary star depletes its core helium at about 7.1\,Myr, producing a $13.2\mso$ BH with a dimensionless spin parameter of 0.148.

The rejuvenated star also undergoes expansion during its post-MS phase, and the rotational angular momentum of the core is efficiently extracted by the envelope, reflected by the increase of surface specific rotational angular momentum. 
At 7.7\,Myr, the rejuvenated star triggers a Case B mass transfer towards the BH, when its outer envelope remains radiative, whereas a single post-main-sequence star normally develops a convective envelope  (Supplementary Fig.\,\ref{kipp:rejuvenated_star_isolation_and_normal_single_star}). 
During the Case B mass transfer phase triggered by the rejuvenated star, the mass donor loses 20$\mso$ material, most of which is ejected out of the binary system, as strong radiation power generated by accreted material prohibits  mass accretion faster than the Eddington accretion rate of the BH accretor. Consequently, the BH accretes negligible angular momentum, and its dimensionless spin parameter stays the birth value. Then, after a core helium burning phase of 0.4\,Myr. The initial secondary star forms a $19.1\mso$ BH with a dimensionless spin parameter of 0.271. The outcome BBH has an orbital period of 1.4\,d, corresponding to a merger timescale of 1.49\,Gyr.

The effective spin parameter of the BBH is 0.22, and both BHs are mildly spinning. Most of the angular momentum of the pre-collapse star is lost during fast expansion of the hydrogen-rich envelope, which later gets stripped by mass transfer, leaving a slow-rotating helium core. However, this angular momentum loss can be avoided in Case A-Case A systems, producing high-spin second-born BHs (Supplementary Section \ref{app:CaseA-CaseA}). 

\subsection{An example model of Case A-Case A systems\label{app:CaseA-CaseA}}

\begin{figure*}[!ht]
    \centering
    \includegraphics[width=\linewidth]{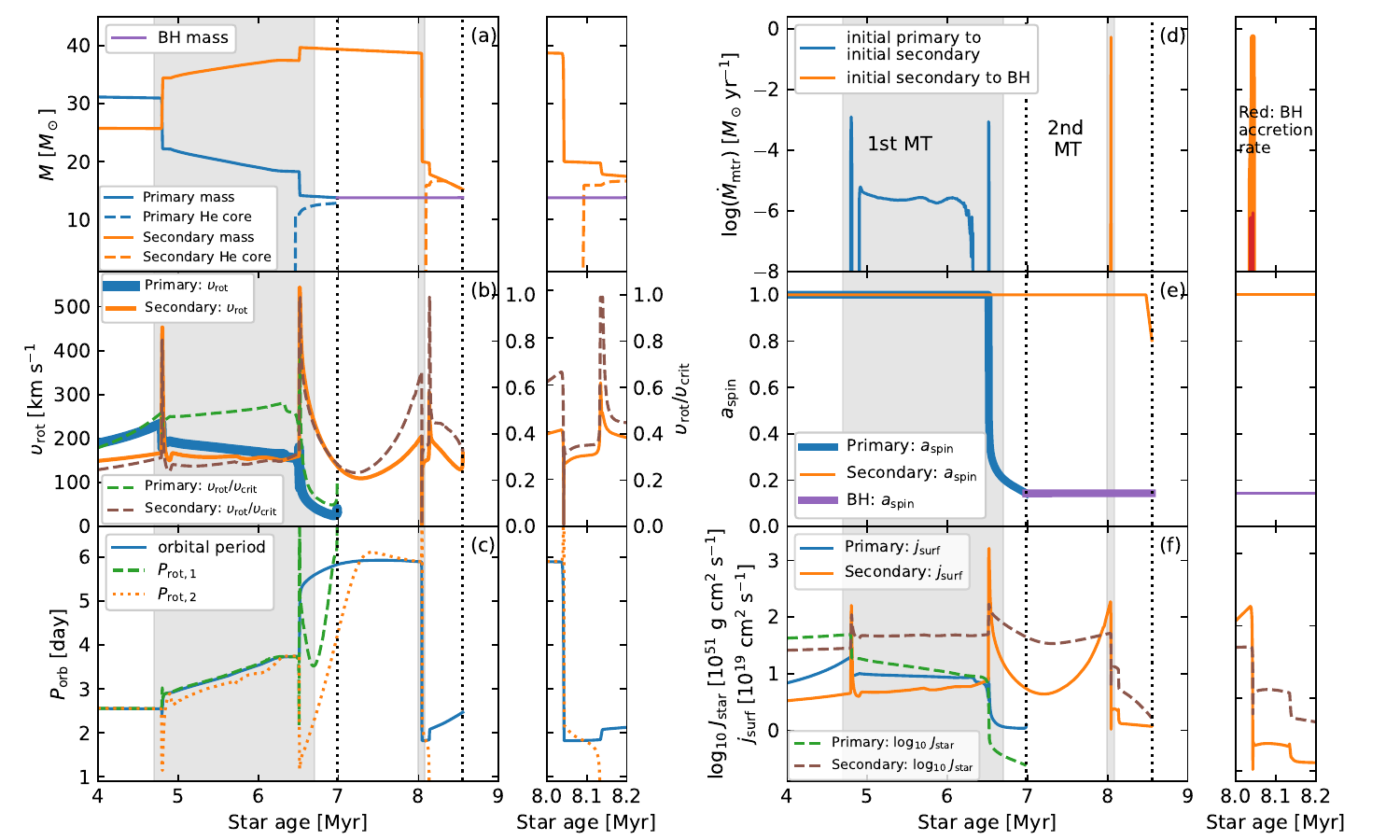}
    \caption{ \textbf{Evolution of a Case A-Case A system.} The depicted model is computed with an initial primary mass of $31.6\mso$, an initial mass ratio of 0.82, and an initial orbital period of 2.5\,d. The lines and colours have the same meaning as Supplementary Fig.\,\ref{kipp:example_model}.  } 
    \label{example_CaseA-CaseA}
\end{figure*}

\begin{figure*}[!ht]
    \centering
    \includegraphics[width=\linewidth]{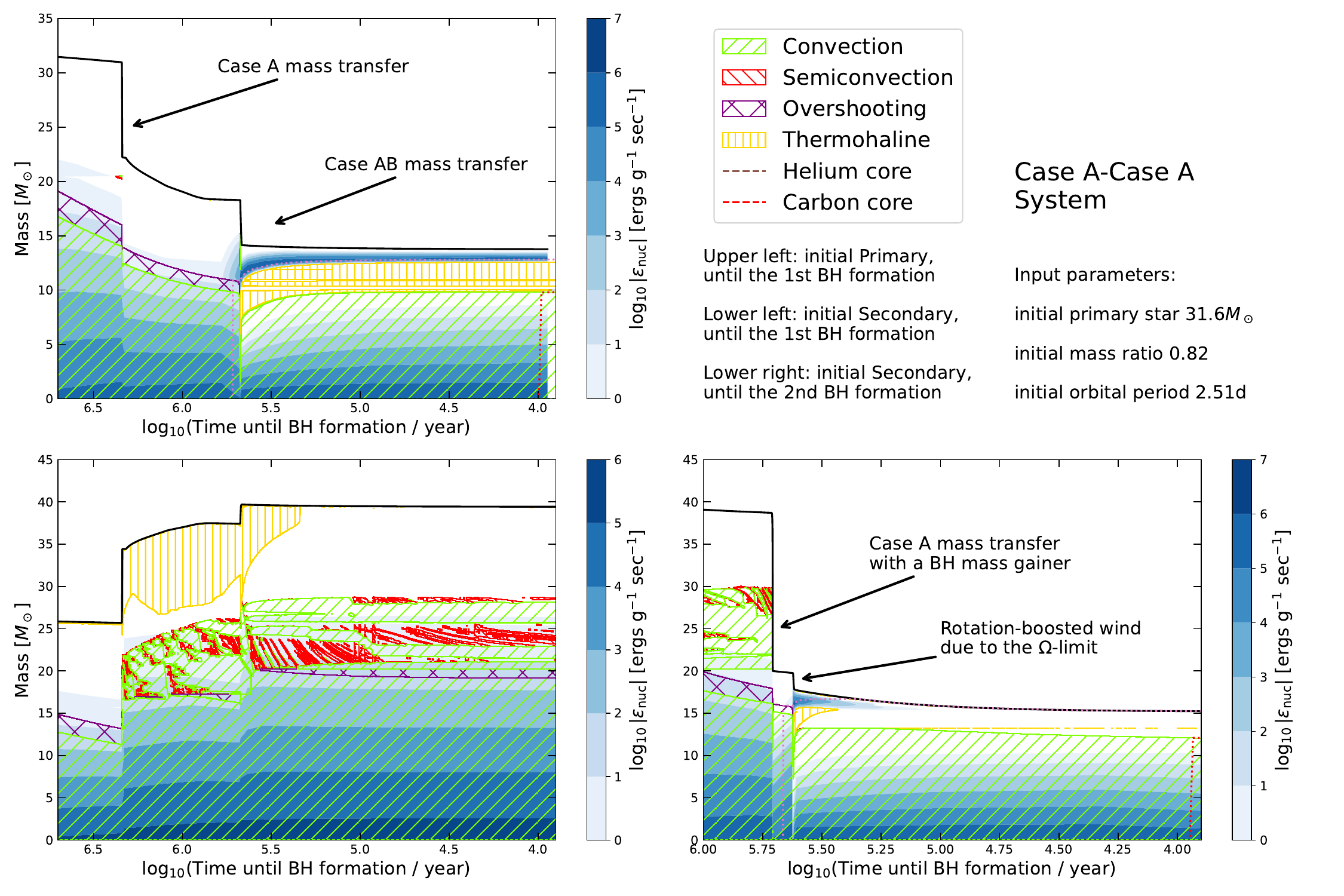}
    \caption{\textbf{Kippenhahn diagrams of the stellar components of the Case A-Case A model presented in Fig.\,\ref{example_CaseA-CaseA}}. The meanings of lines, colours, and hatching patterns are the same as the main text Fig.\,\ref{example_model}.}
    \label{kipp:CaseA-CaseA}
\end{figure*}

In this section, we present a Case A-Case A model, which is computed with $\mi=31.6\mso$,  $\qi=0.82$, and  $\porbi=2.5\,$d (0.4 in log scale). This model produces a merging BBH, containing a high-spin second-born BH.
We present the detailed evolutionary history in Fig.\,\ref{example_CaseA-CaseA} and the Kippenhahn diagrams in Fig.\,\ref{kipp:CaseA-CaseA}.

Similar to the Case A-Case B example model, the first mass transfer phase contains a fast Case A, a slow Case A, and a Case AB phase. The secondary star can be spun up to above 400$\kms$ by tides. During the Case AB phase, it can reach critical rotation. After the first mass transfer phase, the mass gainer is synchronised to the orbit within 0.5\,Myr.
At about 7\,Myr, the primary star forms a $13\mso$ BH. Then, the 39.5$\mso$ rejuvenated star fills its Roche lobe at 8\,Myr with core hydrogen mass fraction dropping to about 0.02. This late Case A mass transfer makes the orbital period shrink from 6\,d to about 2\,d. After about 0.1\,Myr, the donor star depletes its remaining hydrogen in the core. It continues contracting even after the ignition of hydrogen-shell burning 
(Supplementary Section \ref{app:BH-ZAMS-2}), which pushes the star to its $\Omega$-limit (Supplementary Fig.\,\ref{aspin_CaseA-A_Jdot}). This unusual radius evolution makes the system avoids the Case AB phase, and thus the stripped star retains a significant amount of rotational angular momentum, resulting in a 15$\mso$ BH with a spin parameter of about 0.8, which is much higher than previous simulations on the stable mass transfer channel (Supplementary Section \ref{app:BHspin-compare}). 
The birth orbital period of the BBH is about 2.47\,d, corresponding to a merger timescale of 7.7\,Gyr. The structure of the rejuvenated star is also similar to the Case A-Case B model (main text Fig.\,\ref{kipp:example_model}), which develops a off-centre convective zone due to accretion. After its core hydrogen depletion, stellar winds remove the remaining envelope, including the hydrogen-burning shell.

\subsection{Black hole masses\label{app:BH-mass}}

\begin{figure}[!ht]
    \centering
    \includegraphics[width=0.8\linewidth]{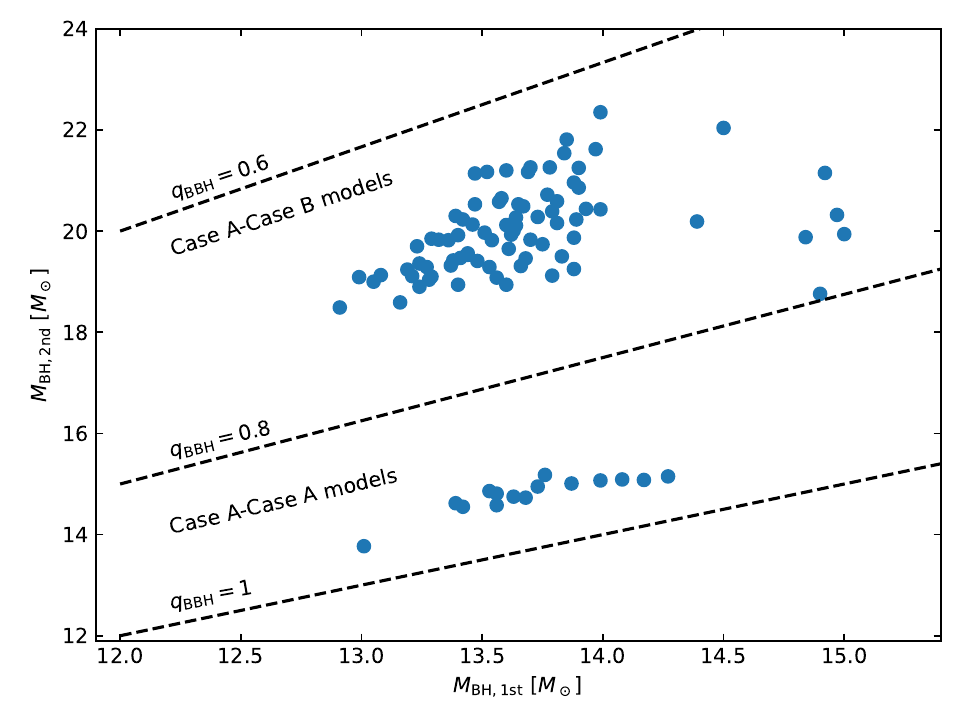}
    \caption{\textbf{Masses of the BH components in the merging BBHs presented in the main text Fig.\,\ref{BH-BH}.} The X- and Y-axes are the masses of the first-born BH, $M_{\rm BH,1st}$, and the second-born BH, $M_{\rm BH,2nd}$, respectively. The black dashed lines indicate equal-$q_{\rm BBH}$ lines with the value of $q_{\rm BBH}$ labelled by the nearby text. The models below/above the $q_{\rm BBH}=0.8$ line correspond to Case A-Case A/Case B types.}
    \label{BH-mass-2}
    \end{figure}

Figure\,\ref{BH-mass-2} presents the masses of both first- and second-born BHs in our merging BBHs. The first-born BHs have masses in 13--14$\mso$, while the second-born BH have masses about 13--23$\mso$. Due to the substantial mass accretion during the first mass transfer, our second-born BHs are always more massive than the first-born BHs. 
The masses of the second-born BHs show a gap at about $16\mso$, which likely arises from the spareness of our grid points.

\subsection{Spin parameters}

\begin{figure}[!ht]
    \centering
    \includegraphics[width=0.9\linewidth]{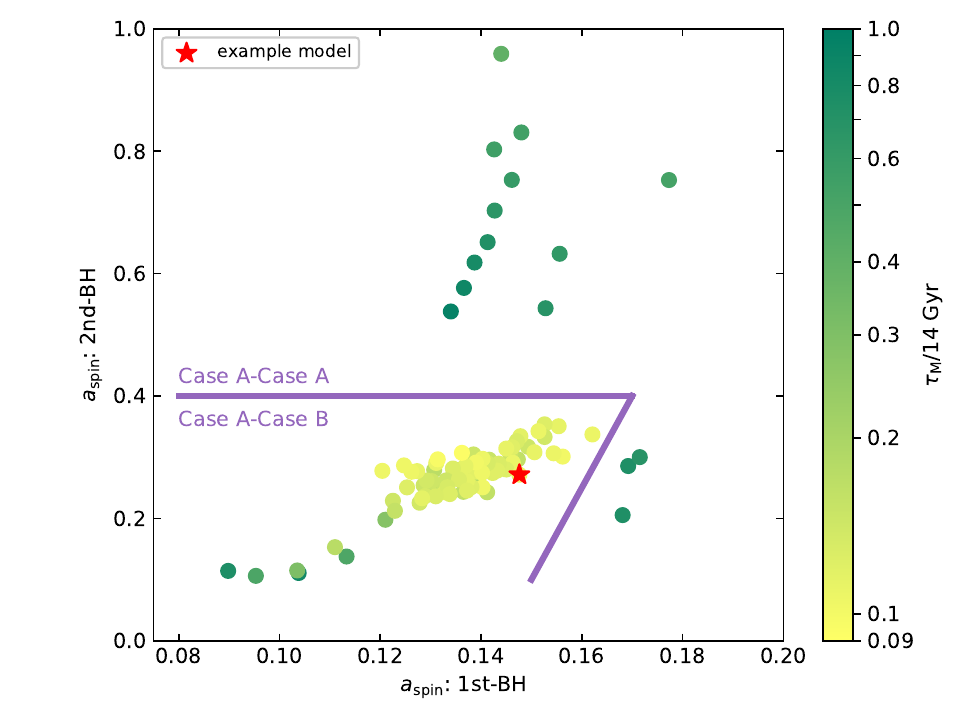}
    \caption{\textbf{Spin parameters of the BH components of our merging BBHs.} The X-axis is the spin parameter of the first-born BH ($a_{\rm spin}$: 1st-BH), and Y-axis corresponds to that of the second-born BH ($a_{\rm spin}$: 2nd-BH). The red star marks the Case A-Case B example model presented in the main text Fig.\,\ref{example_model}. The purple line is the hand-drawn boundary between Case A-Case A and Case A-Case B models for presentation purposes.}
    \label{BBH_aspin}
\end{figure}

Figure\,\ref{BBH_aspin} shows the spin parameters of BH components of our merging BBHs. The first-born BHs have spin parameters below 0.18 in all systems. For the second-born BHs, Case A-Case A systems produce spins of 0.5--1.0, while Case A-Case B systems yield 0.1--0.4. Three Case A-Case A BHs show spins similar to those of Case A-Case B systems, as their rejuvenated donors expand more than in other Case A-Case A models. The discontinuity between our Case A-Case A and -Case B models is likely due to the spareness of our grid points (main text Fig.\,\ref{grid_outcomes}).

\subsection{\texorpdfstring{The $\tau_{\rm M}\text{--}\chi_{\rm eff}$ correlation}{The TauM--Chi correlation}\label{app:tauM-chi}}

\begin{figure}[!ht]
    \centering
    \includegraphics[width=0.8\linewidth]{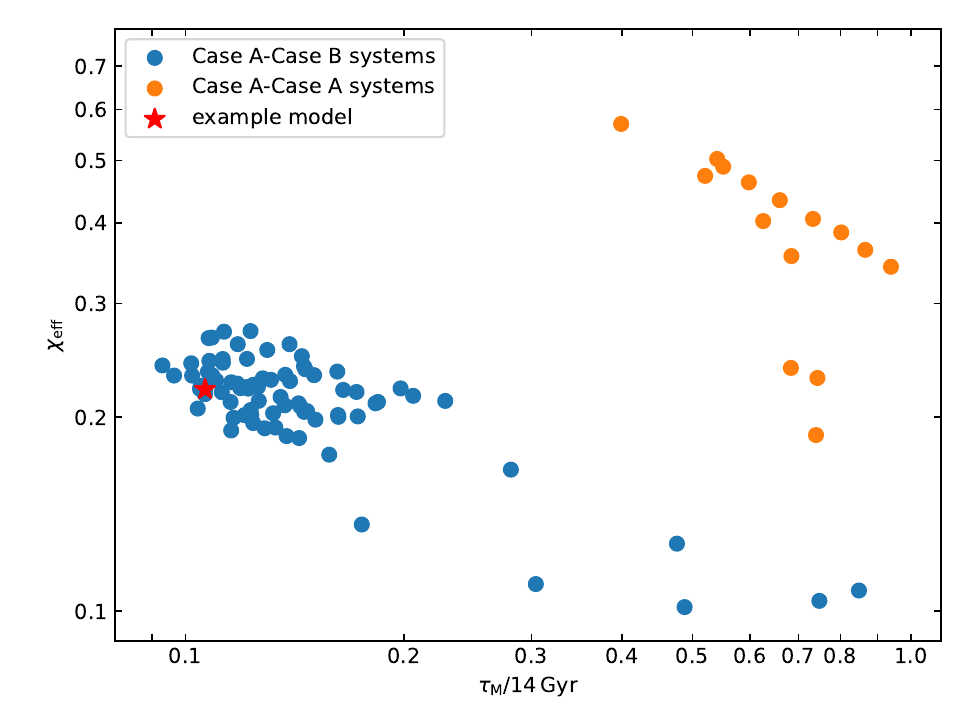}
    \caption{\textbf{Correlation between the merger timescale $\tau_{\rm M}$ and effective spin parameter $\chi_{\rm eff}$.} Blue and orange dots correspond to merging BBHs formed from Case A-Case B and Case A-Case A systems, respectively. The red star marks the Case A-Case B example model presented in the main text. 
    }
    \label{tauM-Xeff}
\end{figure}

Our model expects a correlation between the merger timescale $\tau_{\rm M}$ and the effective spin parameter $\chi_{\rm eff}$ (Fig.\,\ref{tauM-Xeff}). For Case A-Case B systems, merging BBHs with short $\tau_{\rm M}$ trend to have high $\chi_{\rm eff}$. When the merger timescale approaches the Hubble time, 14\,Gyr, the effective spin decreases to 0.1. This correlation reflects the effects of tidal interactions that a longer merger timescale corresponds to a wider BH-Wolf-Rayet star orbit, where tidal spin-up is less efficient, or the Wolf-Rayet star is synchronised to a longer period. Case A-Case A systems show a similar trend. However, unlike Case A-Case B models, the rotational angular momentum of the progenitor of the second-born BH is set by the $\Omega$-limit instead of tidal interaction (Supplementary Section \ref{app:BH-ZAMS-2}), which is therefore more uncertain.  In addition, the minimum merger timescale is affected by the evolutionary stage of the donor star at the onset of the second mass transfer, which is about 2\,Gyr for Case A-Case B systems but 6\,Gyr for Case A-Case A systems. A similar dependence on the mass transfer type is also seen in BH-ZAMS models \cite{Klencki2025arXiv250508860K}.

\section{Comparison with BH-ZAMS models\label{app:BH-ZAMS}}

\subsection{Mass transfer stability\label{app:BH-ZAMS-1}}

\begin{figure*}[!ht]
    \centering
    \includegraphics[width=0.45\linewidth]{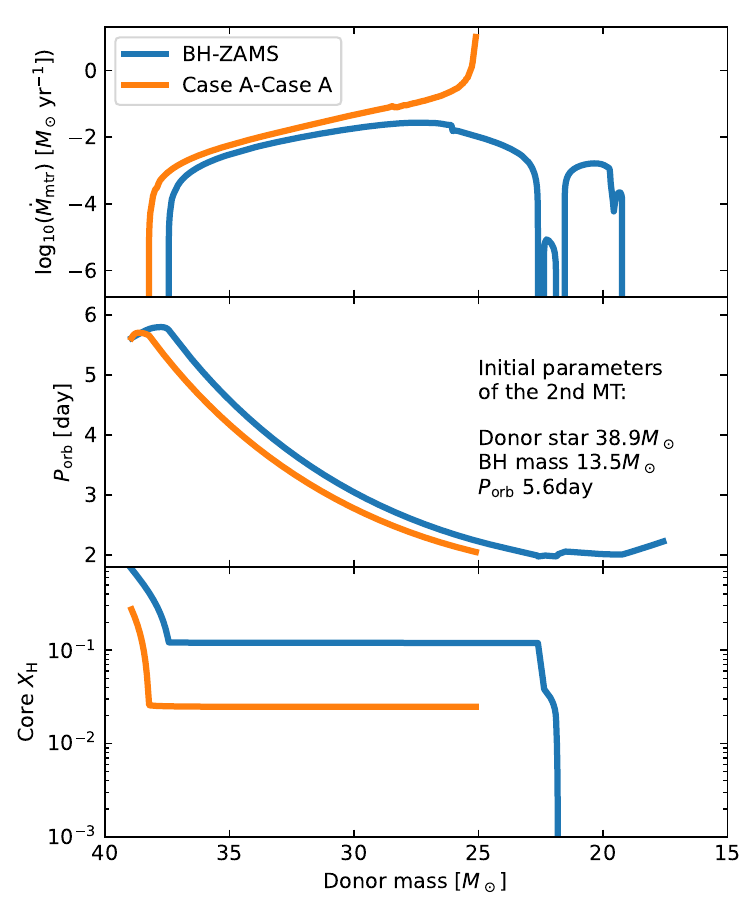}
    \includegraphics[width=0.45\linewidth]{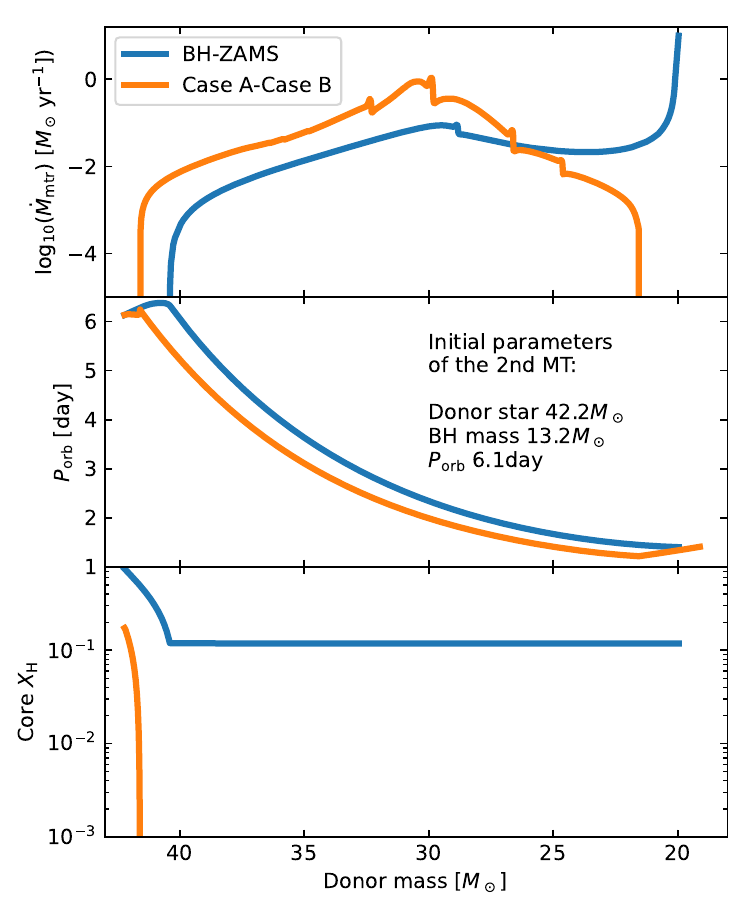}
    \caption{\textbf{Comparison of the second mass transfer of the full-evolution models (orange lines) and the BH-ZAMS models (blue lines) in terms of mass transfer stability.} The left column compares a Case A-Case A model,  $(M_{\rm 1,ZAMS}/\mso,\,q_{\rm ZAMS},\log P_{\rm orb,ZAMS}/\text{day}) = (31.6,\,0.8,\,0.38)$, with its BH-ZAMS counterpart, and the right column compares a Case A-Case B model, $(M_{\rm 1,ZAMS}/\mso,\,q_{\rm ZAMS},\log P_{\rm orb,ZAMS}/\text{day}) = (31.6,\,0.9,\,0.33)$, with its BH-ZAMS counterpart. The panels, from top to bottom, present mass transfer rate $\dot{M}_{\rm mtr}$, orbital period, and core hydrogen mass fraction versus donor mass.}
    \label{Compare_NormalCaseA}
\end{figure*}

In the main text Fig.\,\ref{BH-MS}, we have shown that the mass transfer stability of our Case A-Case A system and Case A-Case B system is very different from that obtained using the models initialised by a ZAMS star and a BH (BH-ZAMS models; e.g., \cite{Marchant2021,Gallegos-Garcia2021,Klencki2025arXiv250508860K}). This is caused by the structural difference of the mass donor. 
Rejuvenated stars develop an off-centre convective zone in their envelope (cf., Supplementary Section\,\ref{app:kipp-single-and-rejuvenated}), which destabilises the reverse mass transfer (see also \cite{Soberman1997,Ge2020,Klencki2025arXiv250508860K}), and rejuvenated stars deplete its core hydrogen at a smaller radius due to the higher surface helium mass fraction (Fig.\,\ref{chemical_strucutre_at_H_dep}), allowing Case B reverse mass transfer to occur in tight binaries.
Consequently, our stable Case A-Case B models cover a region which cannot be reached by even the tightest BH-ZAMS models, while BH-ZAMS models avoid unstable mass transfer at a shorter orbital period than our Case A-Case A systems (Fig.\,\ref{BH-MS}). 
In this section, we further illustrate this point by computing two BH-ZAMS counterpart models of two selected full-evolution models, Model-1 with $(M_{\rm 1,ZAMS}/\mso,\,q_{\rm ZAMS},\log P_{\rm orb,ZAMS}/\text{day}) = (31.6,\,0.8,\,0.38)$, and Model-2 with $(M_{\rm 1,ZAMS}/\mso,\,q_{\rm ZAMS},\log P_{\rm orb,ZAMS}/\text{day}) = (31.6,\,0.9,\,0.33)$, and we compare their evolution in Fig.\,\ref{Compare_NormalCaseA}. The initial masses and orbital periods of the BH-ZAMS models are taken to be the values at the formation time of the first-born BH of the selected full-evolution models.

Model-1 is a Case A-Case A system.
The rejuvenated mass donor fills its Roche lobe at orbital period about 5.7\,d, with its core hydrogen mass fraction dropping to about 0.02-0.03, and the mass transfer becomes unstable, when the Roche lobe reaches the off-centre convective zone at about 25$\mso$, producing a star-BH merger. However, the BH-ZAMS counterpart model survives the Case A mass transfer.

Model-2 is the Case A-Case B example model (cf. Fig.\,\ref{example_model} and Supplementary Section\,\ref{app:kipp-CaseA-CaseB}), producing a merging BBH, whereas the BH-ZAMS counterpart model---with a core hydrogen of 0.1 at the Roche-lobe filling---experiences an unstable Case A mass transfer, and resulting in a star-BH merger. The merger occurs when the remaining stellar mass approaches 20$\mso$, where the Roche lobe meets the boundary of the convective core of the MS mass donor.

\subsection{Spin parameters of the second-born BHs\label{app:BH-ZAMS-2}}

Our Case A-Case A example model produces a second-born BH with much higher spin than the Case A-Case B example model (cf., Supplementary Sections \ref{app:kipp-CaseA-CaseB} and \ref{app:CaseA-CaseA}). As we shown here, this high spin is also unexpected by the BH-ZAMS counterpart model, which is computed with  the same orbital configuration and rotational velocity as our Case A-Case A example model at the formation time of the first-born BH.

\begin{figure}[!ht]
    \centering
    \includegraphics[width=0.9\linewidth]{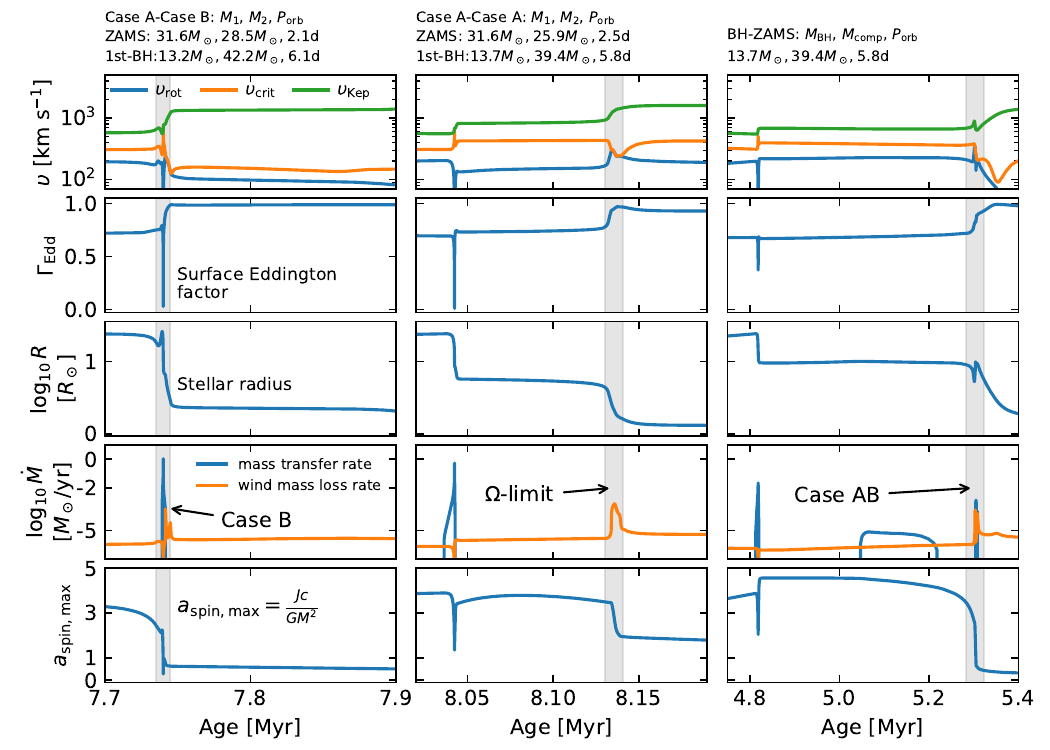}
    \caption{\textbf{Evolution during the second mass transfer  of the Case A-Case B example model (left), the Case A-Case A model (middle), and the BH-ZAMS counterpart model of the Case A-Case A model (right).} The text on the top indicates the types of the models and the parameters at ZAMS and at the first BH formation. The panels, from top to bottom, present the evolution of rotational velocity $\upsilon_{\rm rot}$ (blue) with critical rotational velocity $\upsilon_{\rm crit}$ (orange) and Keplerian rotational velocity $\upsilon_{\rm Kep}$ (green), surface Eddington factor $\Gamma_{\rm edd}$, stellar radius $R$, 
    maximal spin parameter $a_{\rm sin,max}$ (spin parameter but allowed to exceed 1). 
    The grey background outlines Case B mass transfer in the Case A-Case B model, the $\Omega$-limit in the Case A-Case A model, and Case AB mass transfer in the BH-ZAMS model, which are the key phases for rotational angular momentum loss.}
    \label{aspin_CaseA-A_Jdot}
\end{figure}

Figure\,\ref{aspin_CaseA-A_Jdot} compares the evolution during the second mass transfer of the above models.
In the Case A-Case B and BH-ZAMS models, $a_{\rm spin,max}$\footnote{dimensionless spin parameter but allowed to exceed 1.} decreases significantly during the Case B or Case AB mass transfer. The donor stars contract upon core hydrogen depletion but rapidly expand due to hydrogen-shell burning ignition (i.e., the mirror principle). During this expansion, rotational angular momentum is efficiently taken away from the core due to the Spruit-Taylor dynamo. Consequently, the stripping of the envelope leaves a slow-rotating core. 

\begin{figure}[!ht]
    \centering
    \includegraphics[width=\linewidth]{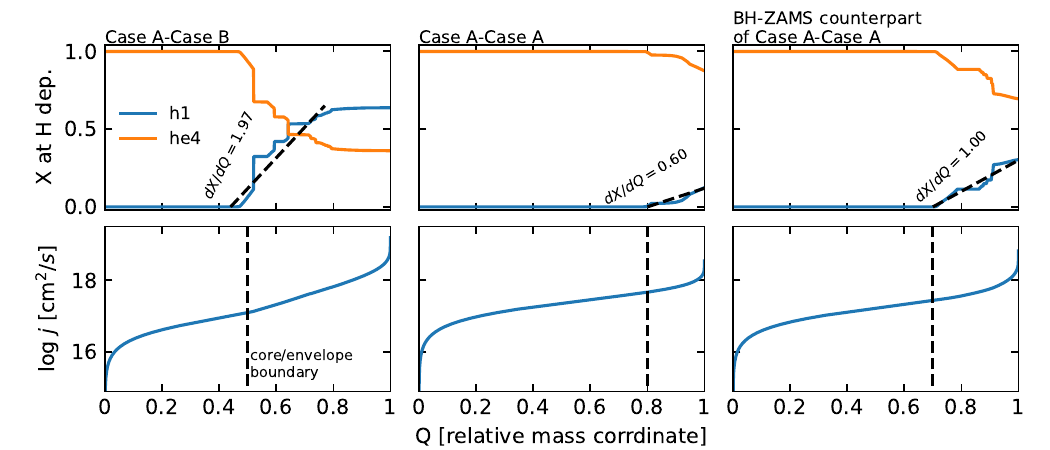}
    \caption{\textbf{Profiles of chemical abundance (upper panel) and specific angular momentum (lower)  at core hydrogen depletion of the donors in the models presented in Fig.\,\ref{aspin_CaseA-A_Jdot}}. The left, middle, and right panels correspond to the Case A-Case B, Case A-Case A, BH-ZAMS counterpart models. The X-axis is relative mass coordinate. In the upper panel, the blue and orange lines represent the mass fractions of hydrogen and helium. In the lower panel, the black dashed line corresponds to the core/envelope boundary.}
    \label{fig:CaseACaseA-Hprofile}
\end{figure}

However, this process does not occur in the Case A-Case A example model. At core hydrogen depletion, the rejuvenated star contracts like the other models. However, the star continues contracting even after the ignition of the hydrogen-burning shell (see Fig.\,\ref{kipp:CaseA-CaseA} for the Kippenhahn diagram). During this contraction, it spins up and becomes brighter, pushing the surface to the $\Omega$-limit (i.e. rotational velocity equal to critical velocity), and stellar wind strips the remaining envelope. 
At core hydrogen depletion, this star's envelope contains less angular momentum than others (Fig\,\ref{fig:CaseACaseA-Hprofile}), and this amount cannot increase later, as the stars avoid substantial expansion due to its small envelope mass (0.2 of the total mass), shallow hydrogen gradient (slop: 0.6), and high envelope helium mass fraction (above 0.8), as shown in Fig\,\ref{fig:CaseACaseA-Hprofile}.  Hence, it preserves a large fraction of its rotational angular momentum until core collapse, producing a fast-rotating second-born BH.

\section{Interpretation of observed black hole spins \label{app:observed-BH-spin}}

\subsection{Spins of black holes in high-mass X-ray binaries}

Measurements of BH spins in X-ray binaries rely on how spin-induced relativistic effects modify the X-ray emission from accretion discs\cite{Remillard2006ARA&A..44...49R}. Currently, dynamically confirmed BH high-mass X-ray binaries, Cygnus X-1 \cite{Miller-Jones2021}, LMC X-1 \cite{Orosz2009}, and M33 X-7 \cite{Ramachandran2022}, all have reported spins close to 1, which, however, is suggested to sensitively depend on assumed accretion disc models \cite{Reynolds2021ARA&A..59..117R,Zdziarski2024ApJ...962..101Z,Zdziarski2024ApJ...967L...9Z}. For example, including a warm comptonisation layer on top of the disc can reduce the inferred spin to below 0.1 \cite{Zdziarski2024ApJ...967L...9Z}.

If the BH spins in high-mass X-ray binaries were truly as high as near 1, they would be in tension with our theoretical expectations (cf., Figs.\,\ref{kipp:example_model} and \ref{example_CaseA-CaseA}). 
The observed high BH spins may be explained by a weak envelope-core coupling \cite{Qin2019}, which, however, still lacks observational supports. It may also be explained that these BHs were significantly spun up by a recent mass transfer episode\cite{Qin2022RAA....22c5023Q,Xing2025A&A...693A..27X}, but this would require a mass accretion rate exceeding those in current general relativistic radiation magnetohydrodynamic numerical simulations of  super-Eddington accretion discs \cite{Xing2025A&A...693A..27X}. 
Therefore, we conclude that the origin of the observed high-spin BHs in wind-fed high-mass X-ray binaries remains unclear.

\subsection{Spins of black holes in merging binary black holes\label{app:BBH-spin}}

GWTC-4 sources likely arise from multiple formation channels\cite{Zevin2021ApJ...910..152Z,LARSEN2025102459,Sridhar2025arXiv251122093S}, and our models reproduce a subset of the observed $\chi_{\rm eff}$ and $q_{\rm BBH}$. In this section, we discuss whether the chemically homogenous evolution (CHE) channel and the common envelope evolution (CEE) channel can explain the observed $\chi_{\rm eff}$ and $q_{\rm BBH}$ with the same BH formation prescription as the main text.

For the CHE channel, the resulting BBHs are expected to be near equal-mass, as stellar components rapidly evolve to equal-mass during the overcontact phase \cite{Marchant2016,Hastings2020A&A...641A..86H,Popa2025arXiv250900154P}. However, the effective spins, $\chi_{\rm eff,CHE}$, can vary widely from 0 to 1 \cite{Marchant2016,Marchant2024A&A...691A.339M} due to wind braking during the Wolf-Rayet star phase. 

For the CEE channel, the first mass transfer is usually expected to be stable \cite{Belczynski2016Natur.534..512B}, which produces a slow-rotating first-born BH \cite{Belczynski2020A&A...636A.104B}. After surviving the following CEE, the tidal spin-up of the BH-helium star binary can produce a second-born BH with $a_{\rm spin}$ potentially up to 1\cite{Belczynski2020A&A...636A.104B,Qin2018A&A...616A..28Q,Bavera2021RNAAS...5..127B}. 
Taking $a_{\rm spin}=1$ for the second-born BH, the maximal effective spin, $\chi_{\rm eff,CEE}$, that can be achieved in the CEE channel is
\begin{equation}
\begin{split}
    \chi_{\rm eff,CEE} & = \frac{M_{\rm BH,1}\,a_{\rm spin,1} + M_{\rm BH,2}\, a_{\rm spin,2}}{M_{\rm BH,1}+M_{\rm BH,2}}= \frac{\,a_{\rm spin,1} + q_{\rm BBH}}{1+q_{\rm BBH}},
    \label{eq:Xeff-CEE}
\end{split}
\end{equation}
where $M_{\rm BH}$ is the BH mass, and subscriptions ``1" and ``2" correspond to the first- and second-born BH respectively. Here the first-born BH is assumed to be the primary BH. However, high accretion efficiencies in wide binaries cannot be excluded theoretically \cite{Shao2014,Vinciguerra2020,Schurmann2025arXiv250323878S,Bao2025arXiv250602662B}, which may lead to mass-ratio reversal in merging BBHs (i.e., the second-born BH becomes more massive than the first-born BH; \cite{Broekgaarden2022ApJ...938...45B}). In this case, the maximal effective spin, $\chi_{\rm eff,CEE,MRR}$, becomes
\begin{equation}
\begin{split}
    \chi_{\rm eff,CEE,MRR} & = \frac{M_{\rm BH,1}\,a_{\rm spin,1} + M_{\rm BH,2}\, a_{\rm spin,2}}{M_{\rm BH,1}+M_{\rm BH,2}}= \frac{ q_{\rm BBH}\,a_{\rm spin,1}+1}{1+q_{\rm BBH}}.
    \label{eq:Xeff-CEE-MRR}
\end{split}
\end{equation}

\begin{figure}[!ht]
    \centering
    \includegraphics[width=0.9\linewidth]{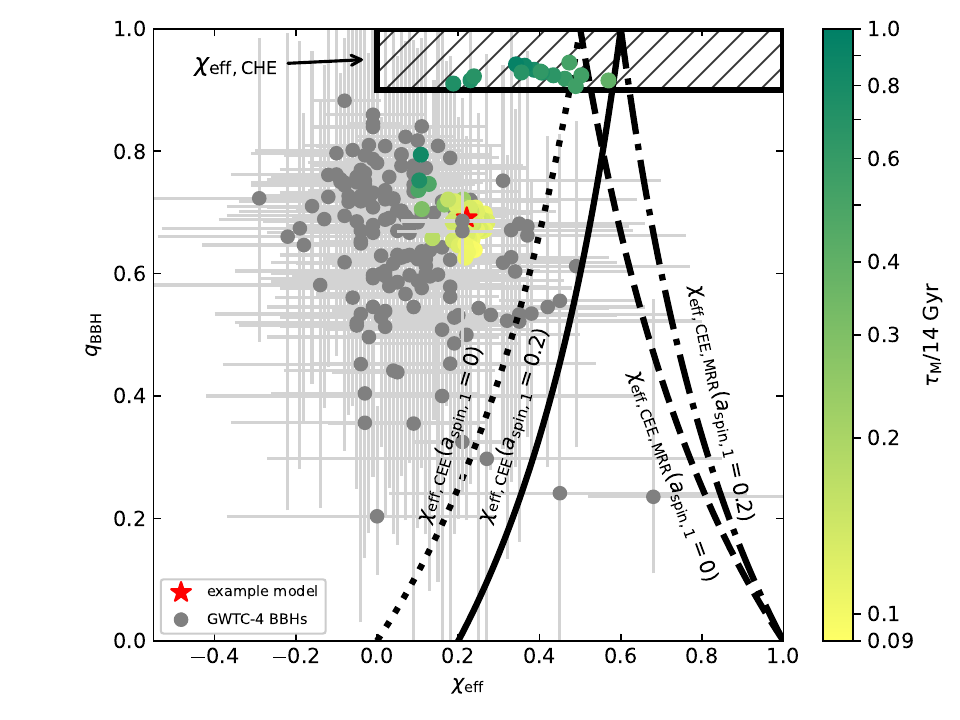}
    \caption{\textbf{Predicted effective spin $\chi_{\rm eff}$ and mass ratio $q_{\rm BBH}$ of merging binary black holes formed through isolated binaries.} The colorbar corresponds to the ratio of the merger timescale to the Hubble time for merging BBHs formed from the stable mass transfer channel. The black lines are the maximal effective spins achievable through the common envelope evolution channel (Eqs.\,\eqref{eq:Xeff-CEE} and \eqref{eq:Xeff-CEE-MRR}, with $a_{\rm spin,1}$ set to 0 and 0.2). The black hatched region outlines the effective spins predicted by the chemically homogenous evolution channel based on ref.\cite{Marchant2016}. The red star marks the Case A-Case B example model (Main text Fig.\,\ref{example_model} and Supplementary Section \ref{app:kipp-CaseA-CaseB}).
    The grey dots represent the observed merging BBHs in the Gravitational-Wave Transient Catalog 4 (GWTC-4; \cite{LVK2025arXiv250818082T}).}.
    \label{fig:BBH-Xeff-CEE}
\end{figure}

In Fig.\,\ref{fig:BBH-Xeff-CEE}, we present the estimated $\chi_{\rm eff,CHE}$, $\chi_{\rm eff,CEE}$ and $\chi_{\rm eff,CEE,MRR}$. We consider two natal spins for the first-born BH, 0 and 0.2 for the CEE channel. The CHE channel occupies only a limited parameter space on the $\chi_{\rm eff}$--$q_{\rm BBH}$ plane, but it may account for the most massive stellar-born BH mergers known so far \cite{Popa2025arXiv250900154P}, e.g., GW231123\cite{Abbott2025arXiv250708219T}.
All observed $\chi_{\rm eff}$  values lie below the theoretical limits, $\chi_{\rm eff,CEE}$ and $\chi_{\rm eff,CEE,MRR}$. 
Hence it not impossible that a large fraction of GWTC-4 sources may be formed through the CEE channel. We note that some GWTC-4 sources show strong evidence for mass-ratio reversal if they were formed in isolated binaries.
GW190403\_051519  ($85_{-33}^{+27.8}\mso+20_{\rm -8.4}^{+26.3}\mso$ with $\chi_{\rm eff}=0.68_{-0.43}^{+0.16}$) and GW200208\_222617 ($51_{-30}^{+103}\mso+20_{\rm -5.5}^{+9.2}\mso$ with $\chi_{\rm eff}=0.68_{-0.46}^{+0.42}$) lie above  $\chi_{\rm eff,CEE}$ but still below $\chi_{\rm eff,CEE,MRR}$. In addition, GW241011 and GW241110 show evidence for high-spin primary BHs \cite{Abac2025ApJ...993L..21A}. This feature is also consistent with the expectation of mass-ratio reversal, although their non-negligible spin–orbit misalignments may indicate a dynamical origin\cite{Abac2025ApJ...993L..21A} or BH spin tossing during collapse \cite{Tauris2022}. BBHs with negative $\chi_{\rm eff}$ can arise from dynamical interactions, which predict isotropic spins \cite{Rodriguez2019PhRvD.100d3027R,Mandel2022PhR...955....1M,Ye2025arXiv250707183Y}, but they may also be produced by isolated binaries if BHs are born with significant momentum kicks \cite{Kalogera2000}, or if BH spins are tossed during core collapse \cite{Tauris2022}.

\section{Comparison of effective spin predictions from previous works and from our full-evolution models\label{app:BHspin-compare}}

Accurate predictions of BH spins require detailed considerations of differential rotation and internal angular momentum transport, both of which are absent in rapid population codes. These studies usually assume first-born BHs have small natal spins, motivated by the effects of the Spruit-Taylor dynamo \cite{Yoon2005A&A...443..643Y,Fuller2019ApJ...881L...1F,Qin2019}. The spin of the second-born BH is affected by the tidal spin-up of its progenitor during the BH-Wolf-Rayet (WR) star phase,  \cite{Qin2018A&A...616A..28Q,Belczynski2020A&A...636A.104B,Bavera2021RNAAS...5..127B,Fuller2022MNRAS.511.3951F}, which is implemented in rapid codes by using fitting formulae based on MESA BH-WR models, as in COMPAS\cite{Bavera2020A&A...635A..97B,Broekgaarden2022ApJ...938...45B}, StarTrack\cite{Belczynski2020A&A...636A.104B,Olejak2021ApJ...921L...2O,Olejak2024A&A...689A.305O}, and COSMIC\cite{Bavera2021A&A...647A.153B,Zevin2022ApJ...933...86Z}. 

With the above method, past rapid population synthesis studies have produced various predictions for the effective spins of merging BBHs formed through the CEE channel. For the stable mass transfer channel, rapid codes typically predict low spins, and often do not expect this channel to be a major contributor to the observed merging BBHs (e.g., \cite{Belczynski2020A&A...636A.104B,Bavera2021A&A...647A.153B}; but see \cite{Olejak2024A&A...689A.305O,Willcox2025arXiv251007573W}). The chemical structure of rejuvenated stars is not captured in these studies, which is likely the reason for the discrepancy between our predictions and those from rapid codes. Recently, a few prescriptions have been developed in order to capture the effects of Case A mass transfer in rapid codes \cite{Schurmann2024A&A...690A.282S,Shikauchi2025ApJ...984..149S,Brcek2025arXiv251213838B}.  It has also been proposed that the first-born BHs may reach high spins if BHs can accrete at super-Eddington rates during stable mass transfer \cite{Bavera2021A&A...647A.153B,Zevin2022ApJ...933...86Z,Shao2022ApJ...930...26S}, while our models assume Eddington-limited BH accretion.
These findings are similar to those  obtained with the POSYDON framework\cite{Bavera2023NatAs...7.1090B,Xing2025A&A...693A..27X}, which does not show the features of Case A-Case A systems either. The reason may lie in the treatment of mass gainers at different evolutionary stages (see \cite{Fragos2023ApJS..264...45F} for detailed description of POSYDON framework) or the difference in metallicities (Our models: SMC metallicity determined by \cite{Brott2011}; POSYDON v1: solar metallicity \cite{Fragos2023ApJS..264...45F}).

\section{Mass ratios of the merging BBHs predicted by the stable mass transfer channel}\label{app:qBBH}

Different evolutionary channels predict distinct mass ratios for merging BBHs. Dynamical interactions allow a wide range of mass ratios \cite{Cook2024arXiv241110590C,Ye2025arXiv250707183Y}, with dense clusters showing a preference for equal-mass BBHs \cite{Ye2025arXiv250707183Y}. For isolated binaries, BBHs formed through the CHE channel have mass ratios above 0.9 (Supplementary Section \ref{app:BBH-spin}), while those through the CEE channel span a wide range depending on assumptions about mass transfer stability (e.g., \cite{Bavera2021A&A...647A.153B,Olejak2021ApJ...921L...2O,Broekgaarden2022ApJ...938...45B}). Meanwhile, GWTC-4 BBHs show a preference at a mass ratio of about 0.74 \cite{LVK2025arXiv250818083T}, a feature suggested to be linked to the stable mass transfer channel \cite{vanSon2022ApJ...940..184V}. Our full-evolution models confirm this trend (main text Fig.\,\ref{BH-BH} and Supplementary Fig.\,\ref{fig:BBH-Xeff-CEE}). 
In this section, using an approach similar to that of ref.\cite{vanSon2022ApJ...940..184V}, we show that this preference in mass ratio arises naturally from mass transfer physics. 

According to our BH formation prescription, the mass of the first-born BH $M_{\rm BH,1st}$ is given by 
\begin{equation}
    M_{\rm BH,1st} = M_{\rm 1,ZAMS} - \Delta M,
\end{equation}
where $\Delta M$ is the mass lost during the first mass transfer phase. We neglect wind mass loss for simplicity. Since we assume an Eddington-limited BH accretion, 
we treat $M_{\rm BH,1st}$ as a constant.
Let $M_{\rm 2,ZAMS}$ denote the ZAMS mass of the secondary star. After the first mass transfer phase, the secondary mass becomes $(M_{\rm 2,ZAMS} + \Delta M\, \varepsilon)$, where $\varepsilon$ is the mass transfer efficiency. Then the mass of the second-born BH $M_{\rm BH,2nd}$ is 
\begin{equation}
    M_{\rm BH,2nd} = f_{\rm BH}(M_{\rm 2,ZAMS}+\Delta M \varepsilon),
\end{equation}
where $f_{\rm BH}$ is the fraction of the secondary mass that finally collapses into the BH. Then we obtain
\begin{equation}
    \frac{M_{\rm BH,1st}}{M_{\rm BH,2nd}} = \frac{1-\Delta M/M_{\rm 1,ZAMS}}{f_{\rm BH}(q_{\rm ZAMS}+\varepsilon\,\Delta M/M_{\rm 1,ZAMS})}.
    \label{eq:qzams-qbbh}
\end{equation}
Our 31.6$\mso$ primary star produces a BH of 13-14$\mso$, implying $\Delta M/M_{\rm 1,ZAMS}\simeq0.6$. The value of $f_{\rm BH}$ is different in Case A-Case A and -Case B models. Our example models give
\begin{equation}
    f_{\rm BH} \simeq
    \begin{cases}
        0.450 ~~~\text{for Case A-Case B model}\\
        0.375 ~~~\text{for Case A-Case A model}
    \end{cases}.
    \label{eq:f-BH}
\end{equation}

\begin{figure}[!ht]
    \centering
    \includegraphics[width=0.9\linewidth]{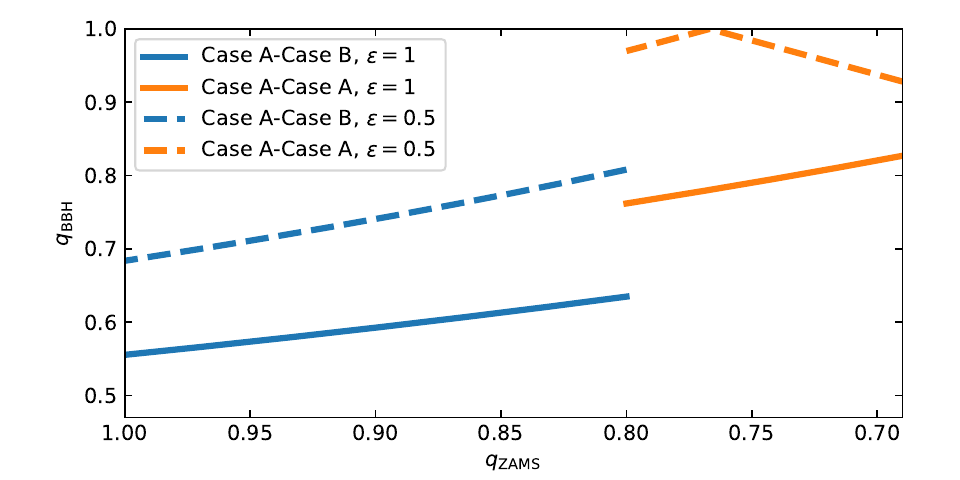}
    \caption{\textbf{Estimated mass ratios $q_{\rm BBH}$ of merging BBHs formed through the stable mass transfer channel as a function of the ZAMS mass ratio $q_{\rm ZAMS}$ using Eq.\,\eqref{eq:qzams-qbbh}.} Blue and orange lines are the values of $q_{\rm BBH}$ estimated for Case A-Case B and Case A-Case A systems. The solid and dashed lines are computed with mass transfer efficiencies $\varepsilon=1$ and $\varepsilon=0.5$.\label{fig:qzams-qbbh}}
    \includegraphics[width=0.9\linewidth]{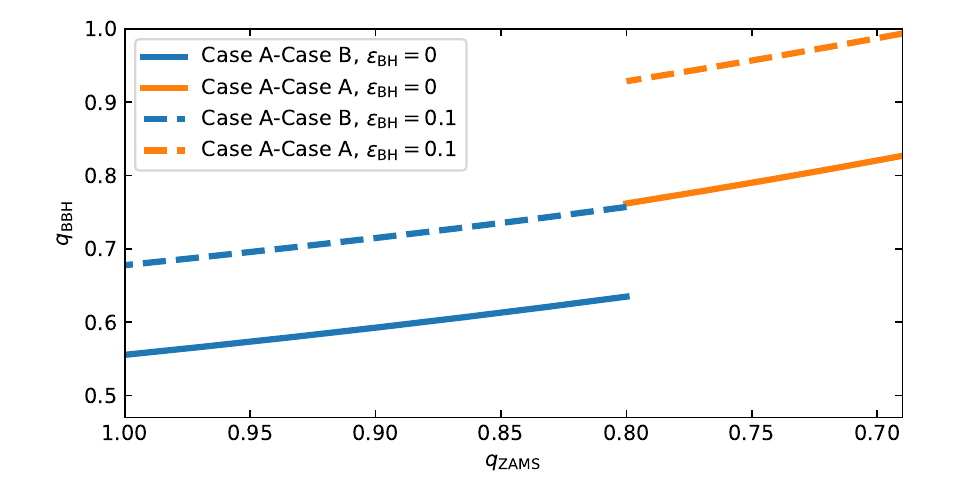}
    \caption{\textbf{Effects of super-Eddington accretion on the mass ratio $q_{\rm BBH}$ of merging BBH formed through the stable mass transfer channel.} Blue and orange lines correspond to Case A-Case B and Case A-Case A binaries with fixed $\varepsilon=1$. The solid and dashed lines are computed with $\varepsilon_{\rm BH}=0$ and $\varepsilon_{\rm BH}=0.1$.\label{fig:qzams-qbbh-2}}
\end{figure}

Using Eq.\,\ref{eq:qzams-qbbh}, we estimate $q_{\rm BBH}$ expected with two mass transfer efficiencies, 1 and 0.5, and we present the results in Fig.\,\ref{fig:qzams-qbbh}. According to the main text Fig.\,\ref{grid_outcomes}, we set the Case A-Case A/B boundary to be $q_{\rm ZAMS}=0.8$. By varying $\varepsilon$ from 0.5 to 1, $q_{\rm BBH}$ varies from 0.7 to 0.55 for Case A-Case B systems, and from 1 to 0.8 for Case A-Case A systems, matching the results of our detailed full-evolution models (main text Fig.\,\ref{BH-BH}). 
The expected $q_{\rm BBH}$ increases as $q_{\rm ZAMS}$ decreases because the secondary star becomes less massive. The sharp transition between Case A-Case A models and Case A-Case B models reflects the change in $f_{\rm BH}$ (Eq.\,\eqref{eq:f-BH}). A lower mass transfer efficiency pushes $q_{\rm BBH}$ upwards, as the second-born BH becomes less massive. In Case A-Case A binaries with $\varepsilon=0.5$, $q_{\rm BBH}$ starts to decline at $q_{\rm ZAMS}<0.76$. This is because that the efficiency is insufficient to reverse the mass ratio in low-$\qi$ binaries. 

In addition, if the first-born BH accretes at a super-Eddington rate, Eq.\,\ref{eq:qzams-qbbh} is modified as 
\begin{equation}
    \frac{M_{\rm BH,1st}}{M_{\rm BH,2nd}} = \frac{1-\Delta M/M_{\rm 1,ZAMS}}{f_{\rm BH}(q_{\rm ZAMS}+\varepsilon\,\Delta M/M_{\rm 1,ZAMS})} + \varepsilon_{\rm BH}\frac{1-f_{\rm BH}}{f_{\rm BH}},
    \label{eq:qzams-qbbh-supEdd}
\end{equation}
where $\varepsilon_{\rm BH}$ is the BH accretion efficiency. Figure \ref{fig:qzams-qbbh-2} presents the $q_{\rm BBH}$ computed by a mild super-Eddington accretion, $\varepsilon_{\rm BH}=0.1$ with $\varepsilon$ fixed to be 1. Compared to the Eddington-limited case, super-Eddington accretion shifts $q_{\rm BBH}$ upwards by 0.1--0.2. 

\end{document}